\documentclass[journal]{IEEEtran}

\usepackage[T1]{fontenc}
\usepackage{cite}
\usepackage{amsmath,amssymb,amsfonts}
\usepackage{algorithm}
\usepackage{algorithmic}
\usepackage{graphicx}
\usepackage{textcomp}
\usepackage{xcolor}
\usepackage{subfigure}
\usepackage{booktabs}
\usepackage{multirow}
\usepackage{diagbox}
\usepackage{url}
\usepackage{threeparttable}
\usepackage{array}
\usepackage{textcase}
\usepackage{rotating}
\usepackage{etoolbox}
\usepackage{caption}
\usepackage{stfloats}
\usepackage{bm}
\usepackage{geometry}
\usepackage{hyperref}

\graphicspath{{figures/}}

\begin{document}
\title{Generative Diffusion Model Driven Massive Random Access in Massive MIMO Systems}

\author{Keke Ying, Zhen Gao, \IEEEmembership{Senior Member,~IEEE}, Sheng Chen, \IEEEmembership{Life Fellow,~IEEE}, Tony Q.S. Quek, \IEEEmembership{Fellow,~IEEE}, \\ and H. Vincent Poor, \IEEEmembership{Life Fellow,~IEEE} 
\thanks{Keke Ying is with the School of Information and Electronics, Beijing Institute of Technology, Beijing 100081, China (e-mail: ykk@bit.edu.cn).}%
\thanks{Zhen Gao (\textit{corresponding author}) is with the State Key Laboratory of CNS/ATM, Beijing 100081, China, also with the MIIT Key Laboratory of Complex-Field Intelligent Sensing, Beijing 100081, China, also with Beijing Institute of Technology (BIT), Zhuhai 519088, China, also with the Advanced Technology Research Institute, BIT, Jinan 250307, China, and also with the Yangtze Delta Region Academy, BIT, Jiaxing 314019, China (e-mail: gaozhen16@bit.edu.cn).} %
\thanks{Sheng Chen is with the School of Computer Science and Technology, Ocean University of China, Qingdao 266100, China (e-mail: sqc@ecs.soton.ac.uk).} %
\thanks{Tony Q. S. Quek is with the Singapore University of Technology and Design, Singapore 487372 (e-mail: tonyquek@sutd.edu.sg).} %
\thanks{H. Vincent Poor is with the Department of Electrical and Computer Engineering, Princeton University, Princeton, NJ 08544 USA (e-mail: poor@princeton.edu).} %
\thanks{The partial of this work has been submitted and accepted to IEEE ICC 2026.}
\vspace*{-5mm}
}

\maketitle

\begin{abstract}
Massive random access is an important technology for achieving ultra-massive connectivity in next-generation wireless communication systems.  It aims to address key challenges during the initial access phase, including active user detection (AUD), channel estimation (CE), and data detection (DD).
This paper examines massive access in massive multiple-input multiple-output (MIMO) systems, where deep learning is used to tackle the challenging AUD, CE, and DD functions. 
First, we introduce a Transformer-AUD scheme tailored for variable pilot-length access. This approach integrates pilot length information and a spatial correlation module into a Transformer-based detector, enabling a single model to generalize across various pilot lengths and antenna numbers.
Next, we propose a generative diffusion model (GDM)-driven iterative CE and DD framework. The GDM employs a score function to capture the posterior distributions of massive MIMO channels and data symbols. Part of the score function is learned from the channel dataset via neural networks, while the remaining score component is derived in a closed form by applying the symbol prior constellation distribution and known transmission model. Utilizing these posterior scores, we design an asynchronous alternating CE and DD framework that employs a predictor-corrector sampling technique to iteratively generate channel estimation and data detection results during the reverse diffusion process. Simulation results demonstrate that our proposed approaches significantly outperform baseline methods with respect to AUD, CE, and DD.
\end{abstract}


\section{Introduction}\label{S1}

\IEEEPARstart{I}n wireless communications, multi-user random access is a critical step in establishing initial communication links. This process involves techniques such as activity user detection (AUD), channel estimation (CE), and multi-user data detection (DD). 
Traditional cellular networks often use scheduling-based access protocols and orthogonal multiple access methods \cite{Ke-IOTJ}. As networks evolve, the 6G vision outlines more stringent performance requirements for scenarios involving massive communication. This vision necessitates a connection density ranging from $10^6$ to $10^8$ devices per square kilometer\cite{IMT-2030}. As the number of users increases, dependence on orthogonal multiple access will lead to excessive overhead, substantially affecting communication efficiency.

To overcome the limitations of existing scheduling-based random access protocols, the grant-free non-orthogonal multiple access scheme has been proposed \cite{Ke-IOTJ}. By eliminating the need for complex handshaking between users and the base station (BS), terminals can directly transmit uplink signals composed of non-orthogonal preamble sequences and payload data. This approach effectively reduces communication delays while enabling more devices to access the network simultaneously. However, it also increases the complexity of designing receiver detection algorithms. Previous studies have proposed several advanced receiver schemes, notably sourced random access and unsourced random access \cite{ngma-mag}. Sourced random access \cite{ll-tsp} employs pre-assigned preamble sequences to distinguish user identities and facilitates user activity detection and channel estimation using these preambles. In contrast, unsourced random access employs a common codebook, utilizing codewords to convey information and select coding, interleaving, and spreading techniques \cite{FASURA}. 
Moreover, reference \cite{wyp_tit} investigates a massive random access scheme utilizing individual user codebooks. This scheme recognizes the uniqueness of user codewords and encodes bits during the codeword selection process, integrating features from both sourced and unsourced random access mechanisms previously described. The reference \cite{wyp_tit} provides an in-depth performance analysis of massive random access under multiple-input multiple-output (MIMO) Rayleigh channels, taking into account the presence or absence of channel state information (CSI) at the receiver and whether the number of active users is known, thereby offering theoretical insights into performance limits. This paper mainly focuses on sourced random access schemes as in \cite{ll-tsp}, tackling algorithm design challenges related to AUD, CE, and DD. For simplicity, we continue to refer to the discussed approach as massive random access.

\subsection{Related Work}\label{S1.1}

For massive random access, identifying the set of active users is a prerequisite for executing subsequent transmission tasks. Studies in \cite{shaoxiaodan-20tsp, chenzhilin-21tsp, cuiying-23ofdm} have developed a covariance-based AUD method.
This approach initially derives the likelihood function of user activity from received signals using a transmission model, framing AUD as a maximum likelihood estimation problem. The coordinate descent method is then employed to iteratively estimate each user's activity. However, the algorithm's computational complexity is high due to the need for computing the inverse of the covariance matrix in each iteration and performing serial iterative estimation for multiple users. To address this challenge, the authors of \cite{aud_transformer, aud_uadformer} proposed advanced neural network-based AUD methods, which adopt binary cross-entropy as the loss function and utilize a heterogeneous Transformer to directly map received signals to the active user support set, achieving superior active user detection performance and low inference latency.

The aforementioned research assumes that the channel between users and the BS is characterized by a multi-antenna Rayleigh channel, and it does not tackle the scenarios of non-Rayleigh channels. In non-Rayleigh environments, a common strategy involves the joint execution of AUD and CE tasks. This is often framed as a compressed sensing (CS) problem, which is solved using traditional greedy algorithms, Bayesian approaches, or deep learning methods. For example, the existing algorithms, such as orthogonal matching pursuit (OMP) \cite{ql-tvt}, approximate message passing (AMP)\cite{kml-tsp, sgl-22twc}, and orthogonal AMP (OAMP) \cite{ykk_jsac}, exploit the inherent sparsity of user activity to recover the user channel matrix while simultaneously identifying active users.
To reduce computational complexity and enhance overall performance, model-driven deep learning approaches \cite{cuiying-21jsac, shaoxiaodan-21jsac} have been developed based on traditional CS algorithm frameworks. By integrating learnable parameters with the conventional backbones, such networks demonstrate superior performance under low pilot overhead conditions while significantly accelerating convergence. However, since the primary aim of CS is CE, with activity detection as a secondary outcome, joint AUD and CE methods typically necessitate higher pilot overhead than covariance-based techniques, which directly target the active user set detection \cite{shaoxiaodan-20tsp}.

Once the AUD and CE processes are completed, the massive MIMO DD problem at BS often becomes overdetermined. Various advanced detection algorithms have been developed to address this challenge. In addition to traditional methods like least squares (LS) and linear minimum mean square error (LMMSE), algorithms such as OAMP \cite{oamp-17access}, expectation propagation (EP) \cite{ep-14tcom}, and deep learning-based techniques \cite{re_mimo-21tsp, oamp-20tsp, gepnet-22jsac, amp_gnn-24twc} have demonstrated excellent performance. However, these DD methods rely on the CSI estimated in the earlier stage, and errors in CSI estimation will degrade the detection performance.

Performing joint CE and DD (JCEDD) can enhance overall achievable performance. The work \cite{oamp-20tsp} presents an iterative JCEDD scheme that accounts for channel estimation errors during data detection and employs a model-driven OAMP unfolding algorithm to improve detection. In the CE phase, demodulated data are used as pilots to augment LMMSE CE, resulting in superior performance. 
Additionally, several studies have addressed the joint AUD, CE, and DD problems, where message-passing algorithms or optimization-based methods have been developed to tackle these tasks \cite{cuiying-jadcedd, kml-blind, sgl-25tsp}. Despite their strong performance, these algorithms require elaborate system configurations and dedicated algorithm design, which may not generalize effectively to practical systems or complex channel environments that are challenging to model, thus limiting their applicability. Consequently, they are not included in the comparisons of this paper. 

In recent years, generative model-based methods have been extensively applied to address inverse problems in communications. Unlike Bayesian algorithms, which require manual assumptions of prior signal distributions, generative models employ neural networks to model the signal prior, offering a more effective representation of complex signal distributions. As detailed in \cite{marius-23ce}, the score matching with Langevin dynamics (SMLD) framework is used to tackle the CE problem. This approach involves estimating the score function, defined as the gradient of a log-probability density function \cite{song-19ncsn}. By combining the channel prior's score function, learned from a neural network, with the likelihood score derived from observations, the posterior estimate of high-dimensional channels can be incrementally sampled from Gaussian noise using annealed Langevin dynamics (ALD) \cite{song-19ncsn,song-20ncsnv2}.
The SMLD framework can also address the MIMO detection problem \cite{nicolas-23twc, nicolas-24tsp, helanxin-24iwcmc}. To handle the non-differentiable symbol constellation prior, a noise-perturbed version of the symbol's distribution serves as a proxy function to derive the score function \cite{nicolas-23twc}. Furthermore, the SMLD framework has been effectively applied to the JCEDD problem \cite{nicolas-24icassp} as well as the quantized CE problem \cite{zxy-arxiv}, demonstrating improved performance over existing deep learning and traditional methods.
 
Another category of generative model frameworks is the denoising diffusion probabilistic model (DDPM) \cite{ddpm}. DDPM methods have been widely applied to denoising tasks in communication systems. For example, the study \cite{fft-dm} proposed a lightweight diffusion model that first obtains an initial CE using LS and then progressively denoises it to enhance CE performance. Additionally, the work \cite{cddm} applied diffusion models to denoise semantic features at the receiver, thereby improving semantic transmission performance.

It has been indicated in \cite{song-21sde} that the DDPM process implicitly calculates score function for each noise level, and both DDPM and SMLD can be unified under the framework of score-based generative models. These models can be described using stochastic differential equations (SDEs), which model both the forward noise-injection and reverse generation processes. This SDE approach offers greater flexibility in sampling algorithms, resulting in improved performance compared to existing generative models.

\subsection{Our Contributions}\label{S1.2}

This paper proposes a new two-stage algorithm to solve the AUD, CE, and DD tasks in massive random access systems. The first stage identifies the active users set (AUS) through a Transformer-based detector, while the second stage utilizes generative diffusion models (GDMs) to perform JCEDD for active user equipment (UEs). The existing Transformer-based deep learning AUD scheme \cite{aud_transformer} fails to effectively address the challenges posed by variable pilot lengths in massive random access problems. Furthermore, the performance of current JCEDD schemes relying on ALD-based generative methods \cite{nicolas-24icassp} has not been comprehensively evaluated and is susceptible to converging to local optima. To tackle these issues, our research makes the following key contributions:

\begin{itemize} 
\item We design a variable pilot length AUD network (VPL-AUDNet) for AUD task. Specifically: 1) We propose the variable pilot length transmission framework and corresponding adaption mechanism that allows a single network to be applicable to different pilot lengths without retraining. 2) We introduce a spatial correlation mechanism to extract spatial correlation features from the received signals, thereby improving network-based AUD performance in complex correlated channels. 

\item We model the CE and DD tasks as reverse diffusion processes that sample from a joint distribution. These
processes gradually transform initial noise into the original channel and data symbols. By employing advanced
predictor-corrector samplers, our scheme outperforms existing JCEDD schemes.

\item To accommodate the distinct characteristics of unknown continuous channel distributions and known discrete data constellation distributions, we propose an asynchronous iterative CE and DD framework. This framework integrates an LMMSE-based CE initialization method, which reduces the number of sampling steps and enhances convergence performance.
\end{itemize}

\begin{figure*}[!b]
\vspace{-4mm}
\centering
\captionsetup{font={footnotesize}}
\includegraphics[width=0.85\textwidth,keepaspectratio]{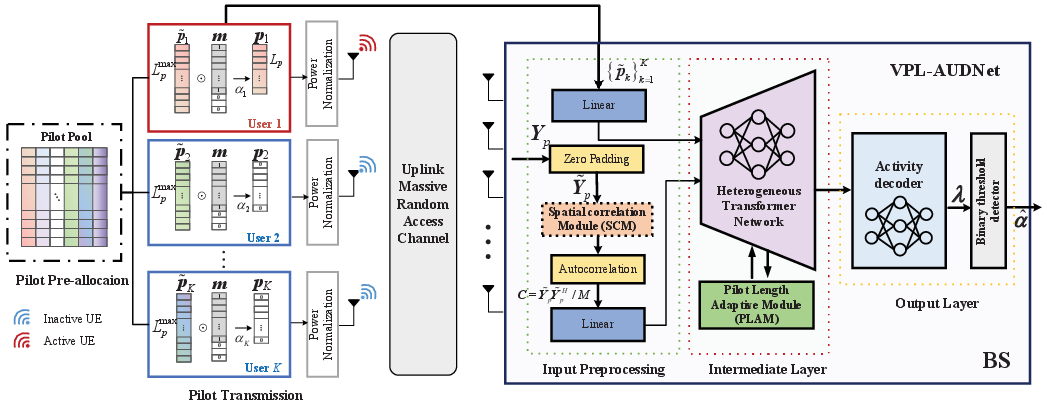}
\vspace{-2mm}
\caption{System diagram of the massive random access with variable-length pilot transmission and corresponding VPL-AUDNet at the receiver.}
\label{fig.aud_diagram} 
\vspace{-1mm}
\end{figure*}

The remainder of this paper is organized as follows. Section~\ref{S2} briefly introduces the massive access system model, and Section~\ref{S3} formulates the problems of AUD and JCEDD. Section~\ref{Sec.AUD} details the proposed Transformer-based AUD scheme, while Section~\ref{Sec.JCEDD} elaborates on our generative models-based JCEDD scheme. Section~\ref{S6} presents the simulation results, and Section~\ref{S7} concludes the paper.

\textit{Notations:} 
Matrices and column vectors are denoted by uppercase and lowercase boldface letters, respectively. $\bm{I}_{N}$ is the $N\times N$ identity matrix, $\bm{0}_{N\times M}$ is the $N\times M$ all-zero matrix, and $\bm{1}_N$ is the $N\times 1$ all-one vector. $(\cdot)^{\rm T}$, $(\cdot)^{*}$, $(\cdot)^{\rm H}$, $(\cdot)^{-1}$, and $\mathbb{E}\{\cdot\}$ denote the transpose, conjugate, Hermitian transpose, inversion, and expectation operations, respectively. $|\mathcal{A}|_c$ indicates the cardinality of the set $\mathcal{A}$. $\mathrm{diag}(\bm{a})$ transforms vector $\bm{a}$ into the corresponding diagonal matrix. $\mathcal{N}\left(\bm{x};\bm{a},\bm{A}\right)$ denotes the probability density function (PDF) of a Gaussian random vector $\bm{x}$ with mean $\bm{a}$ and covariance matrix $\bm{A}$, and the PDF of a complex-valued Gaussian random vector is denoted by $\mathcal{N}_{c}(\cdot;\cdot,\cdot)$.
$\bm{X}_{[:,\mathcal{A}]}$ ($\bm{X}_{[\mathcal{A},:]}$) refers to the sub-matrix comprising the columns (rows) of $\bm{X}$ identified by the index set $\mathcal{A}$, while $\otimes$ and $\odot$ denotes Kronecker and Hadamard product operations, respectively.
The operator $\text{vec}(\cdot )$ stacks a matrix column by column into a vector, and $[\bm{X}]_{i,j}$ is the $(i,j)$-th element of $\bm{X}$. $\left\lfloor \cdot \right\rfloor$ denotes the integer floor operator.
 
\section{System Model}\label{S2}

In this study, we examine a typical narrowband uplink random access scenario where $K$ potential UEs are served by a single BS. The BS is equipped with a uniform planar array (UPA) of $M = M_x \times M_y$ antennas, with $M_x$ and $M_y$ antennas positioned along the $x$-axis and $y$-axis, respectively.
For simplicity, each UE is assumed to use a single antenna. In a specific transmission frame, only $K_a \ll K$ UEs are active due to the sporadic transmission characteristic of massive random access. The activity of a UE is represented by the indicator $\alpha_k$, where $\alpha_k = 1$ indicates that the $k$-th UE is active, and $\alpha_k = 0$ otherwise. By defining the AUS as $\mathcal{A} = \{k \mid \alpha_k = 1, 1 \le k \le K\}$, we have $K_a = |\mathcal{A}|_c$.

For the $k$-th active UE, the uplink frame consists of $L_p$ pilot symbols $\bm{p}_k \in \mathbb{C}^{L_p}$ followed by $L_d$ data symbols $\bm{x}_k \in \mathbb{C}^{L_d}$. These pilot and data symbols form a frame $\bm{s}_k = \left[\bm{p}_k^{\rm T},\, \bm{x}_k^{\rm T}\right]^{\rm T} \in \mathbb{C}^{L}$ of length $L = L_p + L_d$. The channel vector between the $k$-th UE and the BS is denoted by $\bm{h}_k \in \mathbb{C}^{M}$, which is assumed to remain constant throughout the entire transmission frame. The received signal matrix $\bm{Y} \in \mathbb{C}^{L\times M}$ over a frame from all the active UEs is expressed as:
\begin{equation}\label{eqRxS} 
  \bm{Y} = \sum_{k=1}^{K}\alpha_k \bm{s}_k \bm{h}_k^{\rm T} + \bm{W},	
\end{equation}
where $\bm{W} \in \mathbb{C}^{L\times M}$ is the noise matrix, whose elements are symmetric complex additive
Gaussian white noise (AWGN) with zero mean and variance $\sigma_{n}^2$. 

Define $\bm{S}\! \triangleq\! \big[\bm{s}_1, \bm{s}_2, \dots, \bm{s}_K\big]\! \in\! \mathbb{C}^{L\times K}$, $\bm{A}\! \triangleq\! \mathrm{diag}(\bm{\alpha})\! =\! \mathrm{diag}\Big( \big[\alpha_1, \alpha_2, \dots, \alpha_K\big]^{\rm T}\Big)\! \in\! \mathbb{C}^{K\times K}$, $\bar{\bm{H}}\! \triangleq\! \big[\bm{h}_1, \bm{h}_2, \dots, \bm{h}_K\big]^{\rm T}\! \in\! \mathbb{C}^{K\times M}$, and $\bm{H}\! \triangleq\! \bm{A}\bar{\bm{H}}$. Then, the received signal matrix can be expressed as:
\begin{equation}\label{equ.sys1} 
  \bm{Y} = \bm{S}\bm{A}\bar{\bm{H}} + \bm{W} = \bm{S}\bm{H} + \bm{W}.
\end{equation}
By splitting the pilot and data components in $\bm{S}$ into two parts, i.e., $\bm{P}\! \triangleq\! \big[\bm{p}_1, \bm{p}_2, \dots, \bm{p}_K\big]\! \in\! \mathbb{C}^{L_p\times K}$ and $\bm{X}\! \triangleq\! \big[\bm{x}_1, \bm{x}_2, \dots, \bm{x}_K\big]\! \in\! \mathbb{C}^{L_d\times K}$, we have $\bm{S}\! =\! \big[\bm{P}^{\rm T}, \bm{X}^{\rm T}\big]^{\rm T}\! \in\! \mathbb{C}^{L\times K}$ and $\bm{Y}\! =\! \big[\bm{Y}_p^{\rm T}, \bm{Y}_d^{\rm T}\big]^{\rm T}\! \in\! \mathbb{C}^{L\times M}$, where $\bm{Y}_p\! \in\! \mathbb{C}^{L_p \times M}$ and $\bm{Y}_d\! \in\! \mathbb{C}^{L_d \times M}$ correspond to the received pilot and data components, respectively. We also denote $\bm{W}\! = \big[\bm{W}_p^{\rm T}, \bm{W}_d^{\rm T} \big]\! \in\! \mathbb{C}^{L\times M}$ with $\bm{W}_p\! \in\! \mathbb{C}^{L_p \times M}$ and $\bm{W}_d\! \in\! \mathbb{C}^{L_d \times M}$.

\section{Problem Formulation}\label{S3}

The proposed massive random access system is depicted in Fig.~\ref{fig.aud_diagram}, where the BS needs to solve the problems of AUD and JCEDD. Specifically, we divide the AUD, CE, and DD tasks into the two stages. In the first stage, we use a single model to derive the AUS information from the received pilots of variable lengths. In the second stage, we perform the JCEDD using the powerful generation capability of diffusion models. 

\subsection{\textbf{Stage 1-AUD}}\label{stage1} 

Due to activity sparsity, existing CS methods \cite{ql-tvt, kml-tsp, ykk_jsac, cuiying-21jsac, shaoxiaodan-21jsac} recover the entire sparse channel matrix, which requires a large pilot overhead to accurately estimate the CSI. As discussed in Section~\ref{S1}, covariance-based methods \cite{shaoxiaodan-20tsp, chenzhilin-21tsp, cuiying-23ofdm} have not been successfully generalized to non-Rayleigh channels. Additionally, current deep learning-based methods \cite{aud_transformer, aud_uadformer} fail to generalize to variable pilot lengths once the network is trained with a fixed pilot length. To address these issues, we propose a neural network-based method that can generalize to variable pilot lengths and perform well with low pilot overhead in complex correlated channel environments.

We assume that the original pilot sequence for each UE, denoted as $ \tilde{\bm{p}}_k$, is selected from a large pilot pool $\tilde{\bm{P}}\! =\! \left[\tilde{\bm{p}}_1, \tilde{\bm{p}}_2, \dots, \tilde{\bm{p}}_K\right]\! \in\! \mathbb{C}^{L_p^{\text{max}} \times K}$ and is pre-allocated by the BS when the UE initially registers with the BS. The UE identity is then bound to this pilot sequence until it un-registers from the network.
During the transmission procedure as shown in the left part of Fig.~\ref{fig.aud_diagram}, only the first $L_p$ symbols of each pilot sequence are transmitted. That is, each pilot sequence is masked by a binary vector $\bm{m}\! =\! \left[1, \dots, 1, 0, \dots, 0\right]^{\rm T}\! \in\! \left\{0,1\right\}^{L_p^{\text{max}}}$, whose first $L_p$ elements are ones and the remaining $\left(L_p^{\text{max}} - L_p\right)$ elements are zeros. Here, $L_p$ can take a value in the range $[L_p^{\text{min}}, \, L_p^{\text{max}}]$. Let $\bm{M}\! =\! \bm{1}_K^{\rm T}\! \otimes\! \bm{m}\! \in\! \mathbb{B}^{L_p^{\text{max}} \times K}$. Then, at the AUD stage, the BS identifies the active UEs from the received pilot signals, expressed as: 
\begin{equation}\label{equ.aud_mask} 
	\tilde{\bm{Y}}_p = \left(\tilde{\bm{P}}\odot \bm{M}\right)\bm{H} + \tilde{\bm{W}}_p .
\end{equation}
Here, $\tilde{\bm{Y}}_p\! =\! \left[\bm{Y}_p^{\rm T}, \bm{0}_{(L_p^{\text{max}}-L_p)\times M}\right]\! \in\! \mathbb{C}^{L_p^{\text{max}} \times M}$ is the equivalent received pilot after zero-padding it to the maximum pilot sequence length. Similarly, $\tilde{\bm{W}}_p\! =\! \left[\bm{W}_p^{\rm T}, \bm{0}_{(L_p^{\text{max}}-L_p)\times M}\right]\! \in\! \mathbb{C}^{L_p^{\text{max}} \times M}$ is the equivalent noise at the receiver. Since the equivalent channel matrix can be written as $\bm{H}\! =\! \left[\alpha_1\bm{h}_1, \alpha_2\bm{h}_2, \dots, \alpha_K\bm{h}_K\right]^{\rm T}\! \in\! \mathbb{C}^{K \times M}$, estimating the activity $\hat{\bm{\alpha}}\! =\! \left[\hat{\alpha}_1, \hat{\alpha}_2, \dots, \hat{\alpha}_K\right]^{\rm T}\! \in\! \mathbb{B}^{K \times 1}$ is equivalent to recognizing the support set of $\bm{H}$.

The AUD problem can then be formulated as follows: 
\begin{align}\label{eq_problem} 
	\begin{array}{cl}
		\min\limits_{\bm{\Phi}} & \mathbb{E}\left\{\mathcal{L}\big(\bm{\alpha}, \hat{\bm{\alpha}}\big)\right\}, \\ 
		\text{s.t.} & \hat{\bm{\alpha}} = f_{\bm{\Phi}}\left(\tilde{\bm{Y}}_p, \tilde{\bm{P}}, L_p\right), \forall L_p \in \left[L_p^{\text{min}}, L_p^{\text{max}}\right],
	\end{array}
\end{align}
where $f_{\bm{\Phi}}\left(\tilde{\bm{Y}}_p, \tilde{\bm{P}}, L_p\right)$ is a neural network which maps the received signal and pilot sequences to $\hat{\bm{\alpha}}$, $\bm{\Phi}$ represents the parameters of the neural network, and $\mathcal{L}\left(\bm{\alpha}, \hat{\bm{\alpha}}\right)$ is the loss function that quantifies the disparity between $\bm{\alpha}$ and $\hat{\bm{\alpha}}$. 

The advantage of variable pilot-length transmission is not only to avoid training and storing multiple neural networks for different pilot lengths but also to allow the UE to transmit different pilot lengths according to varying traffic densities at different times of the day. This reduces transmission delay while maintaining flexibility. The proposed AUD scheme is detailed Section~\ref{Sec.AUD}.

\subsection{\textbf{Stage 2-JCEDD}}

Given the estimated AUS $\hat{\mathcal{A}}\! =\! \{k\,|\,\hat{\alpha}_k\! =\! 1, 1\! \le\! k\! \le\! K\}$, the  transmission model can be equivalently formulated as: 
\begin{equation}\label{equ.trans_jcedd} 
  \bm{Y} = \bm{S}_{[:,\hat{\mathcal{A}}]} \bm{H}_{[\hat{\mathcal{A}},:]} + \tilde{\bm{W}},
\end{equation}
where $\bm{S}_{[:,\hat{\mathcal{A}}]}\! =\! \Big[\bm{P}_{[:,\hat{\mathcal{A}}]}^{\rm T}, \bm{X}_{[:,\hat{\mathcal{A}}]}^{\rm T} \Big]^{\rm T}\! \in\! \mathcal{X}^{L\times \hat{K}_a}$, with $\mathcal{X}$ denoting the quadrature amplitude modulation (QAM) constellation set, and $\bm{H}_{[\hat{\mathcal{A}},:]}\! \in\! \mathbb{C}^{\hat{K}_a\times M}$. For convenience, define $\bm{S}_a\! \triangleq\! \bm{S}_{[:,\hat{\mathcal{A}}]}, \bm{P}_a\! \triangleq\! \bm{P}_{[:,\hat{\mathcal{A}}]}, \bm{X}_a\! \triangleq\! \bm{X}_{[:,\hat{\mathcal{A}}]}$, 
and $\bm{H}_a\! \triangleq\! \bm{H}_{[\hat{\mathcal{A}},:]}$. Given the known observations $\bm{Y}$ and pilots $\bm{P}_a$, we find the maximum a posteriori estimation of both the data symbols $\bm{X}_a$ and channels $\bm{H}_a$, which can be formulated as:
\begin{equation}\label{equ.joint_post} 
  \big\{\hat{\bm{X}}_a, \hat{\bm{H}}_a\big\} = \arg \max\limits_{\bm{X}_a \in \mathcal{X}^{L_d\times \hat{K}_a}, \bm{H}_a} p\left(\bm{X}_a, \bm{H}_a | \bm{Y}, \bm{P}_a \right).
\end{equation}
The above problem can be divided into the two sub-problems when one of the two variables is assumed to be known:
\begin{subequations}\label{equ.problem} 
	\begin{align}
		\hat{\bm{X}}_a =& \arg \max\limits_{\bm{X}_a \in \mathcal{X}^{L_d\times \hat{K}_a}} p\left(\bm{X}_a| \bm{Y}, \hat{\bm{H}}_a, \bm{P}_a \right), \label{post_Xa_} \\
		\hat{\bm{H}}_a =& \arg ~ \max\limits_{\bm{H}_a} ~ p\left(\bm{H}_a | \bm{Y}, \hat{\bm{X}}_a, \bm{P}_a \right). \label{post_Ha_}
	\end{align}	
\end{subequations} 
Different from typical passing algorithms that factorize the joint distribution (\ref{equ.joint_post}) into a factor graph and then inferring the posterior distribution of each element through message passing, our proposed scheme directly samples the signals from the posterior distributions (\ref{post_Xa_}) and (\ref{post_Ha_}). This approach was first implemented in \cite{marius-23ce, nicolas-23twc, nicolas-24tsp, nicolas-24icassp}, where the ALD sampler is used to sample from the posterior distribution in the CE or DD problem.

The ALD sampler provides a viable method for sampling from complex distributions. Recent advances highlight the significant role of the score function in the ALD sampler, revealing a strong connection to the denoiser in the widely recognized DDPM model. Both concepts can be clarified through the perspective of SDE \cite{song-21sde}. Specifically, sampling from a distribution can be interpreted as solving the reverse SDE of a corresponding forward SDE in the diffusion process. Based on this insight, we address the shortcomings of the existing ALD sampler \cite{nicolas-24icassp} to provide more reliable estimation results for the JCEDD problem. The proposed JCEDD scheme is detailed in Section~\ref{Sec.JCEDD}.

\section{Transformer-Based AUD for Variable Pilot Lengths}\label{Sec.AUD} 

As illustrated in the right part of Fig. \ref{fig.aud_diagram}, the receiver's primary function is to extract correlations between the received signal and the user pilot sequences, facilitating the identification of active users whose pilots contribute to the received signal. We propose a VPL-AUDNet consisting of three main blocks. First, in the preprocessing block, we design a unified method for preprocessing received pilot signals with varying lengths. In the second block, the feature sequences from the previous block are concurrently fed into a neural network for correlation extraction, utilizing a heterogeneous Transformer network as the backbone \cite{aud_transformer} to handle the different physical meanings of the received signal and pilot sequences. To accommodate variable pilot lengths, we incorporate an additional pilot length adaptive module (PLAM) to make the Transformer network cognizant of key information from the input sequence. In the output layer, the activity decoder calculates the correlation between the feature sequences of $\tilde{\bm{Y}}_p$ and the multi-user pilot sequences $\{\tilde{\bm{p}}_k\}_{k=1}^{K}$ to estimate the user activity vector $\bm{\lambda}$. The detailed AUD signal processing flow at the receiver is presented below.

\subsection{Preprocessing}\label{S4.1}

Before inputting the user pilots and received signals into the Transformer-network block, preprocessing is required to ensure dimensional consistency. The user pilots need to be converted into real-valued representations, resulting in $K$ sequences, denoted as $\tilde{\bm{z}}_{k}\! \triangleq\! \big[\Re\{\tilde{\bm{p}}_k\}^{\rm T}, \Im\{\tilde{\bm{p}}_k\}^{\rm T}\big]^{\rm T}\! \in\! \mathbb{R}^{2L_p^{\text{max}}}$, $1\! \le \! k\! \le\! K$, each of length $2L_p^{\text{max}}$. These sequences are then processed through a linear layer to produce the pilot feature sequences $\{\bm{z}_k^{0}\}_{k=1}^{K}$, each of length $d$. Concurrently, to handle varying pilot transmission lengths $L_p$ at the receiver, zero-padding is applied to the received signal $\bm{Y}_p\! \in\! \mathbb{C}^{L_p \times M}$ to convert it into $\tilde{\bm{Y}}_p\! \in\! \mathbb{C}^{L_p^{\text{max}} \times M}$ (see (\ref{equ.aud_mask})). Subsequently, to leverage the multi-antenna observations at the BS, the covariance matrix \cite{aud_transformer, chenzhilin-21tsp} of the received signals is calculated as $\bm{C}\! \triangleq\! \tilde{\bm{Y}}_p \tilde{\bm{Y}}_p^{\rm H}/M\! \in\! \mathbb{C}^{L_p^{\text{max}}\times L_p^{\text{max}}}$. This covariance matrix is transformed into a real-valued vectorized form, denoted as $\tilde{\bm{z}}_{K+1}\! \triangleq\! \big[\Re\{\text{vec}(\bm{C})\}^{\rm T}, \Im\{\text{vec}(\bm{C})\}^{\rm T}\big]^{\rm T}\! \in\! \mathbb{R}^{2\left(L_p^{\text{max}}\right)^2}$. The resulting vector undergoes a linear layer transformation to achieve a dimensionality of $d$, denoted as $\bm{z}_{K+1}^{0}$. Finally, these $K$ pilot sequences and the received signal sequence, $\{\bm{z}_k^{0}\}_{k=1}^{K+1}$, are collectively fed into the subsequent heterogeneous Transformer for further feature extraction.

\begin{figure}[!t]
\centering
\captionsetup{font={footnotesize}}
\includegraphics[width=1\columnwidth]{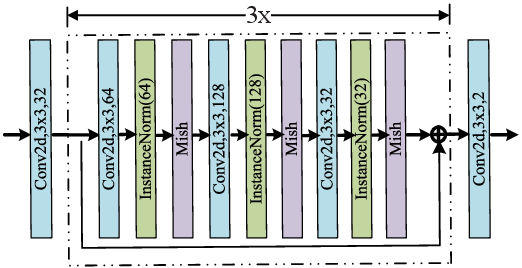}
\vspace*{-1mm}
\caption{Block diagram of the proposed spatial correlation module.}
\label{fig.scm} 
\vspace{-4mm}
\end{figure}

In scenarios involving spatially correlated channels, directly mapping the covariance matrix onto a sequence may fail to effectively capture the antenna correlations. To address this limitation, we introduce a spatial correlation module (SCM) between the zero-padding and autocorrelation processes. Specifically, the received signal $\tilde{Y_p}\! \in\! \mathbb{C}^{L_{\text{max}} \times M}$ is first transformed into a real tensor with dimensions $L_{\text{max}} \times 2 \times M_x \times M_y$ and then inputted into the SCM. To prevent the network parameters from being dependent on the number of antennas, the SCM uses a fully convolutional residual network to extract features along the vertical and horizontal antenna dimensions, as depicted in Fig.~\ref{fig.scm}. The output of the SCM is reshaped to dimensions $L_{\text{max}} \times 2 \times M$ for subsequent autocorrelation calculation. For Rayleigh channels with independent identical distribution (i.i.d.), this module can be omitted. Simulation results in Section~\ref{S6} demonstrate that incorporating the SCM significantly enhances the performance of AUD in correlated channels such as 3GPP channels \cite{38.901}.

\subsection{Heterogeneous Transformer and Activity Decoder}\label{S4.2}

Unlike the traditional Transformer, the heterogeneous Transformer introduced in \cite{aud_transformer} employs distinct sets of weight matrices for input embedding, multi-head attention (MHA), and the feed-forward network (FFN) to account for the different physical meanings of pilot sequences and received signals. To facilitate understanding, we illustrate the heterogeneous Transformer structure and MHA in Appendix \ref{apd1}. Since the relative positional order of different sequences is not meaningful in our AUD task, we do not apply positional encoding as in conventional Transformers. 

Assuming that the heterogeneous Transformer network consists of $N$ encoder layers, we conduct activity estimation for $K$ users based on the $K\! +\! 1$ sequences output by the $N$-th encoder layer. This task is performed by the activity decoder, which computes the heterogeneous MHA between the final output feature sequence of the received signal $\bm{z}_{K+1}^{N+1}$ and pilot sequences $\{\bm{z}_{k}^{N+1}\}_{k=1}^{K}$.
The resulting MHA output, $\boldsymbol{z}_c$, represents the synthesized contributions of all pilot sequences to the received signal.  Next, the user activity is estimated as the weighted correlation between $\bm{z}_c$ and each pilot feature sequence. It is given by:
\begin{equation}\label{eqWC} 
  \lambda_{k}^{\prime} = C \tanh\left(\frac{\bm{z}_c^{\rm T}\bm{W}^o\bm{z}_k^{N+1}}{\sqrt{d}}\right) , 
\lambda_{k} = \frac{1}{1+e^{-\lambda_{k}^{\prime}}}, \forall k. 
\end{equation}
Here, $\{\lambda_k\}_{k=1}^{K}$ is the output of the activity decoder, $\bm{W}^o\! \in\! \mathbb{C}^{d \times d}$ is a learnable parameter matrix, and $C$ is a hyperparamter empirically set to $C\! =\! 10$. User activity is then determined by a binary threshold detector: $\hat{\alpha}_{k}\! =\! \mathbb{I}(\lambda_{k} > 0.5), \forall k$, where the indicator $\mathbb{I}(x)\! =\! 1$ if $x$ is true, and 0 otherwise.

\subsection{Pilot Length Adaptive Module}\label{S4.3}

In variable pilot-length transmission, the input sequence at the receiver undergoes zero-padding, causing some elements to carry no useful information. As a result, the significance of different elements in the input sequence varies. Allowing the neural network to recognize this variability enables adaptive optimization of network parameters based on the corresponding pilot lengths.

\begin{figure}[!t]
\centering
\captionsetup{font={footnotesize}}
\includegraphics[width=1\columnwidth]{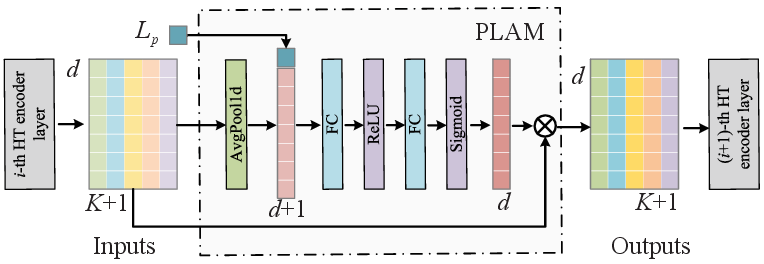}
\vspace*{-5mm}
\caption{Block diagram of the proposed pilot length adaptive module.}
\label{fig.plam} 
\vspace{-4mm}
\end{figure}

Inspired by the attention feature module in \cite{adjscc_tcsvt}, we propose a PLAM. By inserting a PLAM module between each pair of adjacent Transformer encoder layers, we provide additional pilot length information to the receiver. The structure of the proposed PLAM module is illustrated in Fig.~\ref{fig.plam}. Its input consists of $(K\! +\! 1)$ sequences from the output of the previous Transformer encoder layer, and its output comprises 
$(K\! +\! 1)$ sequences of the same dimension, which serve as the input to the next layer.
First, an average pooling operation is applied to the 
$(K\! +\! 1)$ sequences along the sequence length dimension, producing a sequence of length $d$. This $d$-dimensional sequence is then concatenated with the pilot length information $L_p$ and processed through a two-layer fully connected network, generating a $d$-dimensional weight sequence. Finally, this weight sequence is used for position-wise dot multiplication with all input sequences, redistributing the weights of the input feature sequence and allowing the receiver to recognize the varying importance of different elements in the inputs.

\subsection{Loss Function}\label{S4.4}

We use binary cross-entropy as the loss function. The final loss is computed as the average of the loss values across different pilot lengths, given by:
\begin{align}\label{eqLF} 
  \mathcal{L} =& \frac{1}{\vert \mathcal{D} \vert K}\sum_{i=1}^{\vert \mathcal{D} \vert} \sum_{k=1}^{K} \bigg(\alpha_k^{\left(i\right)}\log\lambda_{k}^{\left(i\right)} \nonumber \\
	& \hspace*{10mm}+ \left(1-\alpha_k^{\left(i\right)}\right)\left(1-\log\lambda_{k}^{\left(i\right)}\right) \bigg).
\end{align}
Here, $\mathcal{D}$ denotes the dataset, which consists of training samples $\left\{{\boldsymbol{Y}}_p^{\left(i\right)}, \tilde{\boldsymbol{P}}^{\left(i\right)}, \boldsymbol{\alpha}^{\left(i\right)}\right\}$ with different pilot lengths $L_p$, where $L_p \in \left[L_p^{\text{min}}, L_p^{\text{max}}\right]$.

\section{Generative Diffusion Models Driven JCEDD Using Predictor-Corrector Sampler}\label{Sec.JCEDD} 

Given the active UE set, we need to recover the channel and corresponding data symbols of the active UEs. Before introducing the proposed JCEDD schemes, we briefly review the relationship between existing generative models and their SDE descriptions. We then address the CE problem, detailing the calculation of posterior score functions within the predictor-corrector (PC) sampler framework. Subsequently, we extend the application of the PC sampler to the DD problem. Finally, we combine the CE and DD methods to propose an iterative algorithm for JCEDD and analyze its complexity.

\subsection{Preliminary: Diffusion Process and SDE}\label{sec.sde_review} 

Existing diffusion-based generative models involve corrupting data with multiple noise scales to form a known prior distribution. These models then learn to transform this prior distribution back into the data distribution by gradually removing noise. SDE considers a continuum of distributions that evolve over time according to a diffusion process. The forward diffusion process $\bm{x}(t)$, indexed by the continuous time variable $t\! \in\! (0,\, T]$, progressively diffuses a data point $\bm{x}_0\! \in\! \mathbb{R}^d$ sampled from a dataset with an unknown distribution $p_0(\bm{x})\! =\! p_{\text{data}}(\bm{x})$ into random noise $\bm{x}(T)$ with distribution $p_T(\bm{x})\! =\! \mathcal{N}(\bm{x}; \bm{0}, \sigma_{T}^2 \bm{I})$. For simplicity in notation, we define $p_t(\bm{x})\! \triangleq\! p(\bm{x}(t))$. This process can be described by an SDE that has no trainable parameters \cite{song-21sde, generative_tut}, as follows:
\begin{equation}\label{equ.ori_sde} 
	\mathrm{d}\bm{x} = \bm{f}(\bm{x}, t) \mathrm{d}t + g(t) \mathrm{d} \bm{w},
\end{equation}
where $\bm{f}(\bm{x}, t)\!\!: \mathbb{R}^d\! \times\! \mathbb{R}\! \rightarrow\! \mathbb{R}^d $ is a vector function called the drift coefficient, and the scalar function $g(t)\!\! : \mathbb{R}\! \rightarrow\! \mathbb{R}$ is the diffusion coefficient. Additionally, $\bm{w}$ is a standard Wiener process, defined as $\mathrm{d}\bm{w} = \bm{z}(t)\mathrm{d}t$, where $\bm{z}(t)\! \sim\! \mathcal{N}(\bm{z}(t); \bm{0}, \bm{I})$.

In the reverse process, starting from a sample $\bm{x}_T\! \sim\! p_T(\bm{x})$, random noise is smoothly transformed into a data sample $\bm{x}_0\! \sim\! p_0(\bm{x})$. This reverse process satisfies a time-reversed SDE, which can be derived from the forward SDE given the score function $\nabla_{\bm{x}}\log p_t(\bm{x})$ for the marginal probability of $\bm{x}$ as a function of time, i.e., 
\begin{equation}\label{equ.rsde0} 
	\mathrm{d}\bm{x} = \big(\bm{f}(\bm{x},t) - g(t)^2\nabla_{\bm{x}}\log p_t(\bm{x})\big)\mathrm{d}t + g(t)\mathrm{d}\bar{\bm{w}}.
\end{equation}
Here, $p_t(\bm{x})$ denotes the probability of $\bm{x}$ at time $t$, and $\bar{\bm{w}}$ denotes the Wiener process when time flows backward from $T$ to $0$.
Existing generative models like SMLD \cite{song-19ncsn} and DDPM \cite{ddpm} can be considered specific numerical discretization solvers of the SDE. For instance, in SMLD, the coefficients are $\bm{f}(\bm{x}, t)\! =\! \bm{0}$ and $g(t)\! =\! \sqrt{\frac{\mathrm{d}(\sigma^2(t))}{\mathrm{d}t}}$, with $\sigma(t)$ serving as a predefined noise scheduling function. For DDPM, the drift and diffusion coefficients are given by $\bm{f}(\bm{x}, t)\! =\! -\frac{\sigma(t)\bm{x}}{2}$ and $g(t)\! =\! \sqrt{\sigma(t)}$, respectively. By applying the discrete form of both the forward and reverse SDEs, we can derive the corresponding iteration equations used in SMLD and DDPM.
	
\subsection{Channel Estimation}\label{S5.2}

As discussed in \cite{song-21sde}, the generation process involves solving the reversed SDE given in Eq.~(\ref{equ.rsde0}), where the unconditional generation from this equation yields samples that adhere to the prior probability distribution $p_{0}(\bm{x})$. However, considering Eq.~(\ref{equ.trans_jcedd}) in the JCEDD problem, since the BS receives signal $\bm{Y}$,  we need to consider the conditional generation process where channel should be drawn from the posterior distribution $p\left(\bm{H}_a|\bm{Y}, \hat{{\bm{X}}}_a, {\bm{P}}_a\right)$ to produce the CE result that align with the constraint $\bm{Y}\! =\! \bm{S}_a\bm{H}_a + \bm{W}$.

Specifically, we assume that the data symbols $\hat{\bm{X}}_a$ are known and represent the initial posterior distribution as $p_0\left(\bm{H}_a|\bm{Y}\right)\! \triangleq\! p\left(\bm{H}_a(0)|\bm{Y}\right)$, where $\bm{H}_a(0)\! =\! \bm{H}_a$ represents the ground truth channel to be estimated. The conditional distribution of $\bm{H}_a(t)$ at time $t$ is expressed as $p_t\left(\bm{H}_a|\bm{Y}\right)\! \triangleq\! p\left(\bm{H}_a(t)|\bm{Y}\right)$. According to \cite{Anderson}, the conditional generation process involves sampling from $p_0\left(\boldsymbol{H}_a|\boldsymbol{Y}\right)$, which is equivalent to solving a reverse SDE represented as follows:
\begin{align}\label{equ.rsde1} 
	\mathrm{d}\bm{H}_a =& \big(\bm{f}(\bm{H}_a,t) - g(t)^2\nabla_{\bm{H}_a}\log p_t(\bm{H}_a\vert \bm{Y})\big) \mathrm{d}t \nonumber \\
	& + g(t)\mathrm{d}\bar{\bm{W}}.
\end{align}
Due to the equivalence between DDPM and SMLD during the inference process \cite{kawar_ddrm}, the remainder of this paper will examine the SDE in the SMLD form. Discussions on DDPM are similar and are therefore omitted.

To solve the reversed SDE (\ref{equ.rsde1}), it is essential to derive the posterior score function $\nabla_{\bm{H}_a}\log p_t\left(\bm{H}_a\vert \bm{Y}\right)$. By applying Bayes' theorem, the posterior score function is given by:
\begin{align}\label{equ.scoreH} 
	\nabla_{\bm{H}_a}\log p_t\left(\bm{H}_a\vert \bm{Y}\right) =& \nabla_{\bm{H}_a}\log p_t\left(\bm{Y} \vert \bm{H}_a \right) \nonumber \\
	&+ \nabla_{\bm{H}_a}\log p_t\left(\bm{H}_a\right),
\end{align}
where the \textit{likelihood score function} $\nabla_{\bm{H}_a}\log p_t\left(\bm{Y} \vert \bm{H}_a \right)$ is determined by the signal transmission model (\ref{equ.trans_jcedd}), and the \textit{prior score function} $\nabla_{\bm{H}_a}\log p_t\left(\bm{H}_a\right)$ is determined by the prior distribution of the channel dataset. 

\subsubsection{Calculation of Prior Score Function}\label{sec.prior_score} 

Denote $\bm{h}_k\! \triangleq\! \left[\bm{H}_a\right]_{\left[k,:\right]}^{\rm T}\! \in\! \mathbb{C}^{M},\, \forall k\! \in\! \mathcal{A}$. Since the prior distributions of each active UE's channel are assumed to be independent of each other, we have $\log p_t\left(\bm{H}_a\right)\! =\! \log \prod_{k=1}^{K_a} p_t(\bm{h}_k)\! =\! \sum_{k=1}^{K_a} \log p_t(\bm{h}_k)$. Therefore, the prior score function can be calculated for each UE separately, i.e., 
\begin{align}\label{equ.prior_score} 
  \nabla_{\bm{H}_a}\log p_t\left(\bm{H}_a\right) =& \Big[\nabla_{\bm{h}_1} \log p_t\left(\bm{h}_1\right), \nabla_{\bm{h}_2} \log p_t\left(\bm{h}_2\right),\cdots, \nonumber \\
	& ~ \nabla_{\bm{h}_{K_a}}\log p_t\left(\bm{h}_{K_a}\right)\Big]^{\rm T}.
\end{align}
Therefore, when all the UEs' channels are generated from the same dataset distribution $\bm{p}_{\text{data}}(\bm{h})$, we only need to focus on the prior score function of a single UE.

For simplicity, omit the UE index $k$ and denote $\bm{p}_{0}(\bm{h})\! \triangleq\! \bm{p}_{\text{data}}(\bm{h})$. Our objective is to derive the score function $\nabla_{\bm{h}}\log p_t\left(\bm{h}\right)$, where $p_t\left(\bm{h}\right)$ is the data distribution diffused by noise at time $t$. For complex channel distributions, the prior score function lacks a closed-form expression. Fortunately, the study \cite{Pascal_dsm} has shown that using a neural network to learn the score function $\nabla_{\bm{h}}\log p_t\left(\bm{h}\right)$ is equivalent to denoising score matching (DSM). This can be achieved by training a time-dependent neural network $\bm{s}_{\bm{\theta}}(\bm{h}(t), t)$ using the following loss function:
\begin{equation}\label{eqDSM-lf} 
	\mathcal{L}_{s}\! =\!  \mathbb{E}_t\! \left\{\! \lambda(t) \mathbb{E}\! \left\{\! \left\lVert \bm{s}_{\bm{\theta}}(\bm{h}(t),t)\! -\! \nabla_{\bm{h}(t)} \log p(\bm{h}(t)|\bm{h}(0)) \right\rVert_2^2\! \right\}\! \right\}\! ,
\end{equation}
where the inner expectation is taken over the joint distribution $p(\bm{h}(t), \bm{h}(0))$ with $\bm{h}(0)$ sampled from the dataset distribution $ p_0(\bm{h})$, and the time $t$ is uniformly sampled from $(0,\, 1]$ with $T\! =\! 1$, while $\lambda(t)\! \propto\! 1/\mathbb{E} [ \|\nabla_{\bm{h}(t)} \log p(\bm{h}(t)|\bm{h}(0)) \|_2^2]$ is a positive weight function, and $\bm{\theta}$ collects the neural network parameters. 

Specifically, $ p(\bm{h}(t) | \bm{h}(0))$ denotes the transition kernel from $\bm{h}(0)$ to $\bm{h}(t)$, typically defined by a forward noise diffusion process given by Eq.~(\ref{equ.ori_sde}). For a typical generation model in the SMLD form \cite{song-21sde}, with a known drift coefficient $\bm{f}(\bm{h}, t)\! =\! \bm{0}$ and a diffusion coefficient $g(t)\! =\! \sqrt{\frac{\mathrm{d}(\sigma^2(t))}{\mathrm{d}t}}$, the transition probability function from the original data point $\bm{h}(0)$ to $\bm{h}(t)$ can be derived using \textit{Ito's Lemma} \cite{generative_tut}. This is given by $p\left(\bm{h}\left(t\right)\vert\bm{h}\left(0\right)\right)\! =\! \mathcal{N}\left(\bm{h}(t); \bm{h}(0), \left(\sigma^2(t)\! -\! \sigma^2(0)\right)\bm{I}\right)\! \approx\! \mathcal{N}\left(\bm{h}(t); \bm{h}(0), \sigma^2(t)\bm{I}\right)$,
where $\sigma (t)\! =\! \sigma_{\min}\Big(\frac{\sigma_{\max}}{\sigma_{\min}}\Big)^{t}$ is a predefined noise scheduling function, and the approximation holds when the initial noise variance $\sigma^2(0)\! =\! \sigma_{\min}\! \approx\! 0$. Consequently, we have $\nabla_{\bm{h}(t)} \log p(\bm{h}(t)|\bm{h}(0))\! =\! - \frac{\bm{h}(t)\! -\! \bm{h}(0)}{\sigma^2(t)}$. The final loss function can thus be simplified as:
\begin{equation} 
	\mathcal{L}_{s} =  \mathbb{E}_t \left\{ \lambda(t) \mathbb{E} \left( \left\lVert \bm{s}_{\bm{\theta}}(\bm{h}(t),t) + \frac{\bm{h}(t)-\bm{h}(0)}{\sigma^2(t)} \right\rVert_2^2 \right) \right\},
\end{equation}
which is equivalent to training a denoising neural network that can recover the scaled noise superposed on the channel. With sufficient data and model capacity, DSM ensures that the optimal solution $\bm{s}_{\bm{\theta}^*}(\bm{h}(t), t)$ approximates $\nabla_{\bm{h}} \log p_t(\boldsymbol{h})$ for almost all $\bm{h}(t)$ and $t$. 

Given the powerful expressive capability of the noise conditional score network (NCSN++) \cite{song-21sde} in image generation tasks, and considering that the channel can be viewed as a special kind of image, we use NCSN++ to construct $\bm{s}_{\bm{\theta}^{*}}(\bm{h}(t), t)$. To adapt the channel dimension to the input of NCSN++, the original $M$-dimensional channel vector sample is transformed into a real representation and reshaped into dimensions $2 \times M_x \times M_y$. NCSN++ then takes the channel of these dimensions and a time step index $t$ as inputs. The network outputs the learned channel prior score at time $t$. It should be noted that during the prior score function learning, no information on the pilot symbols, data symbols, received signal, or transmission noise variance is involved in the training process. This means that once the neural network is trained, it can be applied to any dimension of the JCEDD problem as long as the channel distribution remains unchanged.

One special case where the channel prior score function does not need to be learned is when the channel distribution has a known closed form, for example, $\boldsymbol{h}(0)\sim \mathcal{N}\left(\boldsymbol{h}(0); \boldsymbol{\mu}_h, \boldsymbol{R}_{hh}\right)$. Then, the prior score function of the channel in this case can be derived by Tweedie's formula \cite{Tweedie}
	\begin{equation}\label{equ.prior_score_rayleigh}
		\begin{aligned}
			\nabla_{\boldsymbol{h}(t)} \log p(\boldsymbol{h}(t)) &=  \frac{\mathbb{E}\left\{\boldsymbol{h}(0)|\boldsymbol{h}(t)\right\}-\boldsymbol{h}(t)}{\sigma^2(t)}\\
			&= -\left(\boldsymbol{R}_{hh}+\sigma(t)^2\boldsymbol{I}\right)^{-1}\left(\boldsymbol{h}(t)-\boldsymbol{\mu}_h\right),
		\end{aligned}
	\end{equation}
	where the posterior expectation can be derived according to Bayes linear Gaussian model, which is given by
	$\mathbb{E}\left\{\boldsymbol{h}(0)|\boldsymbol{h}(t)\right\} = \boldsymbol{\mu}_h +  \boldsymbol{R}_{hh}\left(\boldsymbol{R}_{hh} +  \sigma^2(t)\boldsymbol{I}\right)^{-1}\left(\boldsymbol{h}(t)-\boldsymbol{\mu}_h\right)$.
	
\subsubsection{Calculation of Likelihood Score Function}\label{sec.likelihood_score} 

We now consider how to derive $\nabla_{\bm{H}_a}\log p_t\left(\bm{Y} \vert \bm{H}_a \right)$.
Given that the channels are viewed as three-dimensional real tensors in prior score learning, we can also transform the transmission model (\ref{equ.trans_jcedd}) into its corresponding real-valued representation.
For simplicity, we retain the notation $\bm{Y}\! =\! \bm{S}_a \bm{H}_a + \bm{W}$ throughout this paper. 
Note that in the SMLD framework, $p\left(\bm{H}_a(t)\vert\bm{H}_a(0)\right)\! \approx\! \mathcal{N}\left(\bm{H}_a(t); \bm{H}_a(0), \sigma^2(t)\bm{I}\right)$. Consequently, the transmission model can equivalently be expressed as:
\begin{equation}\label{eqETM} 
  \bm{Y} = \bm{S}_a\left(\bm{H}_a(t)+\sigma(t)^2\bm{Z}\right) + \bm{W},
\end{equation}
where $\bm{Z}\! \sim\! \mathcal{N}\left(\bm{Z};\bm{0},\bm{I}\right)$ is the diffusion noise superimposed on the original channel $\bm{H}_a(0)$. Therefore, the corresponding transition probability is given by:
\begin{equation}\label{eqTP} 
  p_t\left(\bm{Y} \vert \bm{H}_a \right) \sim \mathcal{N}\left(\bm{Y}; \bm{S}_a\bm{H}_a(t), \sigma_n^2\bm{I} + \sigma^2(t)\bm{S}_a\bm{S}_a^{\rm T}\right).
\end{equation}
Taking the gradient of the log-likelihood, we obtain:
\begin{equation}\label{equ.ll_score} 
  \nabla_{\bm{H}_a}\!\!\log p_t\left(\bm{Y} \vert \bm{H}_a \right)\! =\! \bm{S}_a^{\rm T}\! {\left(\sigma_n^2\bm{I}\! +\! \sigma^2(t)\bm{S}_a\bm{S}_a^{\rm T}\right)\!}^{-1}\!\!\left(\! \bm{Y}\!\!-\!\!\bm{S}_a\bm{H}_a(t)\! \right)\! .
\end{equation}
In the considered JCEDD problem, $\bm{S}_a\! =\! \left[\bm{P}_a^{\rm T},\, \hat{\bm{X}}_a^{\rm T}\right]^{\rm T}$ contains both the pilot symbol ${\bm{P}}_a$ and estimated data symbol $\hat{\bm{X}}_a$ in the iteration process, and the estimated data symbol can enhance the CE. When no data symbol is used, the problem degrades to the conventional pure pilot-aided CE problem.

\begin{figure*}[!b]
\vspace{-5mm}
\centering
\captionsetup{font={footnotesize}}
	\includegraphics[width=0.8\textwidth]{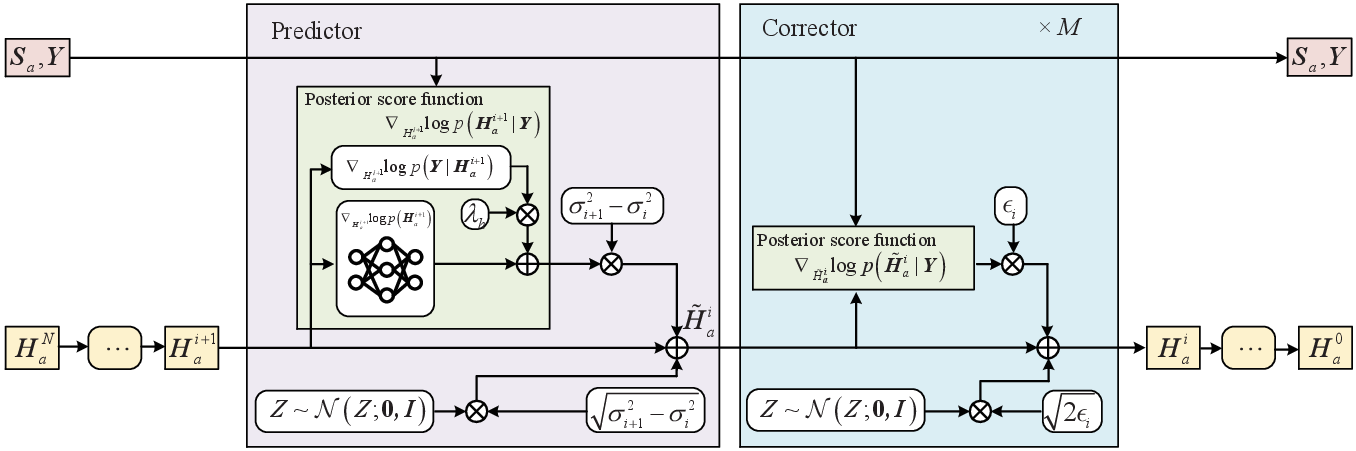}
\caption{Block diagram of the proposed PC sampler-based CE algorithm.}
\label{fig:pc_sampler} 
\end{figure*}

\subsubsection{Channel Sampling Process}\label{S5.2.3} 

Utilizing the derived prior score function and likelihood score function, we numerically solve the reverse SDE (\ref{equ.rsde1}). Specifically, by dividing the interval $t\! \in\! (0,\, 1]$ into $N$ discrete steps, indexed as $i\! =\! 0,1, \dots, N-1$, we obtain the following iterative rule for the reverse diffusion sampler (which is also called \textit{predictor}):
\begin{align}\label{equ.rsampler} 
	{\bm{H}}_a^{i} =& \bm{H}_a^{i+1}\! -\! \bm{f}_{i+1}(\bm{H}_a^{i+1})\! +\! g_{i+1}^2\nabla_{\bm{H}_a^{i+1}}\log p(\bm{H}_a^{i+1}\vert \bm{Y})\nonumber \\
	& + g_{i+1}\bm{Z},
\end{align}
where ${\bm{H}}_a^{i}\! =\! {\bm{H}}_a(t)\vert_{t = \frac{i}{N}}$, $\bm{f}_{i}(\bm{H}_a^{i})\! =\! \bm{f}(\bm{H}_a(t),t)\vert_{t = \frac{i}{N}}$, $g_{i}\! =\! g(t)\vert_{t = \frac{i}{N}}$, and $\bm{Z}\! \sim\! \mathcal{N}(\bm{Z};\bm{0},\bm{I})$. Given the known drift coefficient $\bm{f}(\bm{H}_a, t)\! =\! \bm{0}$ and diffusion coefficient $g(t)\! =\! \sqrt{\frac{\mathrm{d}(\sigma^2(t))}{\mathrm{d}t}}$ in SMLD, Eq.~(\ref{equ.rsampler}) is equivalent to: 
\begin{align}\label{equ.predictor} 
	\tilde{\bm{H}}_a^{i} =& \bm{H}_a^{i+1} + \left(\sigma_{i+1}^2 - \sigma_{i}^2\right) \nabla_{\bm{H}_a^{i+1}} \log p(\bm{H}_a^{i+1}|\bm{Y}) \nonumber \\
	& + \sqrt{\sigma_{i+1}^2 - \sigma_{i}^2}\bm{Z},
\end{align} 
where $\sigma_i\! =\! \sigma(t)\vert_{t = \frac{i}{N}}$. Then, starting from a random initialization $\bm{H}_a^{N}\! \sim\! \mathcal{N}(\bm{H}_a^{N}; \bm{0}, \sigma_{\max}^2 \bm{I})$, the iteration process for the SMLD-based predictor is shown in Fig.~\ref{fig:pc_sampler}. The corresponding discretized form of the score function is given as:
\begin{subequations}\label{equ.post_score_dis} 
	\begin{align}
		& \nabla_{\bm{H}_a^{i}}\! \log p\! \left(\bm{H}_a^{i}\right)\! =\!\! {\Big[\bm{s}_{\bm{\theta}^{*}}\Big(\bm{h}_1^{i}, \frac{i}{N}\Big), \dots, \bm{s}_{\bm{\theta}^{*}}\Big(\bm{h}_{K_a}^{i}, \frac{i}{N}\Big)\! \Big]\!}^{\rm T}\!\! ,\! \\
		& \nabla_{\bm{H}_a^{i}} \log p\left(\bm{Y} \vert \bm{H}_a^{i}\right) = \bm{V}\bm{\Sigma}^{\rm T} \left(\sigma_n^2 \bm{I} + \sigma_i^2\bm{\Sigma} \bm{\Sigma}^{\rm T} \right)^{-1} \nonumber \\
		& \hspace*{33mm} \times \left(\bm{U}^{\rm T} \bm{Y} - \bm{\Sigma}\bm{V}^{\rm T} \bm{H}_a^{i}\right) , \\
		& \nabla_{\bm{H}_a^{i}} \log p\left(\bm{H}_a^{i}\vert \bm{Y}\right) = \lambda_{h}\nabla_{\bm{H}_a^{i}} \log p\left(\bm{Y} \vert \bm{H}_a^{i}\right) \nonumber \\
		& \hspace*{33mm} + \nabla_{\bm{H}_a^{i}}\log p\left(\bm{H}_a^{i}\right).
	\end{align}	
\end{subequations}
In Eq.\,(\ref{equ.post_score_dis}b), we apply singular value decomposition (SVD) to $\bm{S}_a\! =\! \bm{U}\bm{\Sigma}\bm{V}^{\rm T}$ within Eq.\,(\ref{equ.ll_score}) to avoid non-diagonal matrix inversions, thereby reducing computational complexity. Additionally, a scaling parameter $\lambda_h\! =\! 2.5$, as suggested in \cite{mxm_qcs}, is empirically chosen to balance the effects of the likelihood score and prior score, enhancing convergence performance.
 
However, the numerical predictor in Eq.\,(\ref{equ.predictor}) may introduce approximation errors when solving the continuous SDE in a discrete manner. To correct the errors, a score-based Markov Chain Monte Carlo (MCMC) method is used as the \textit{corrector}. This method corrects the marginal distribution of the estimated sample during iterations. A well-known MCMC approach is the ALD, which employs the following iterative process:
\begin{equation}\label{eqMCMC} 
  \bm{H}_a^{i} = \tilde{\bm{H}}_a^{i} + \epsilon_{i} \nabla_{\tilde{\bm{H}}_a^{i}} \log p(\tilde{\bm{H}}_a^{i}|\bm{Y}) + \sqrt{2\epsilon_i} \bm{Z},
\end{equation}
where $\epsilon_{i}\! =\! 2\sigma_{i}\big(r\|\bm{Z}\|_2/\|\nabla_{\tilde{\bm{H}}_a^{i}} \log p(\tilde{\bm{H}}_a^{i}|\bm{Y})\|_2\big)^2$ is the step size in the $i$-th iteration of the ALD algorithm \cite{song-21sde}. Additionally, $r$ is an empirically chosen hyperparameter, set to $r = 0.3$ in our simulations.
Existing ALD-based studies on CE \cite{marius-23ce}, DD \cite{nicolas-23twc}, and JCEDD \cite{nicolas-24icassp} can be regarded as corrector-only samplers with potential for performance improvement. Therefore, we address the JCEDD problem using the PC sampler illustrated in Fig.~\ref{fig:pc_sampler}.

\subsubsection{Overall CE Algorithm}\label{S5.2.4} 

In practical implementations, each iteration of the PC sampler includes one prediction step and $Q$ correction steps, where $Q$ ranges from 1 to 3. The complete CE algorithm is listed as Algorithm~\ref{alg:pc_ce}. Compared to traditional Bayesian algorithms, our method derives prior distribution knowledge through network learning, which avoids the mismatches associated with manually specified prior distributions. Moreover, unlike purely data-driven or model-driven deep learning algorithms based on supervised learning, this method does not require training separate networks for different numbers of active UEs and varying pilot lengths, offering improved generalization and interpretability.

\begin{algorithm}[t]
\caption{PC Sampler-Based Channel Estimation}
\label{alg:pc_ce} 
\begin{algorithmic}[1] 
	\REQUIRE ${\bm{S}_a}$, ${\bm{Y}}$, $\sigma_n^2$, pre-trained denoising network $\bm{s}_{\bm{\theta}^{*}}$, step number $N$, noise scheduling function $\sigma(t)$, and hyper-parameter $\lambda_{h}$; 
	\STATE \textbf{Initialize:} $\bm{H}_{a}^{N}\sim \mathcal{N}\big(\bm{H}_{a}^{N};\bm{0},\sigma_{N}^2 \bm{I}\big)$;
	\STATE \textbf{Compute:} $\bm{S}_a = \bm{U}\bm{\Sigma}\boldsymbol{V}^{\rm T}$;	
	\FOR{$i=N$ to $1$}
		\STATE Calculate $\nabla_{\bm{H}_a^{i}}\log p\left(\bm{H}_a^{i}\vert\bm{Y}\right)$ via (\ref{equ.post_score_dis});
		\STATE Draw $\bm{Z}\sim  \mathcal{N}(\bm{Z}; \bm{0},\bm{I})$;
		\STATE ${\bm{H}}_{a}^{i-1}=\bm{H}_a^{i} + \left(\sigma_{i}^2-\sigma_{i-1}^2\right)\nabla_{\bm{H}_a^{i}} \log p(\bm{H}_a^{i}|\bm{Y}) + \sqrt{\sigma_{i}^2-\sigma_{i-1}^2}\bm{Z}$;  \% \textit{Predictor: line 6}
		\FOR{$j=1$ to $Q$}
			\STATE Draw $\bm{Z}\sim  \mathcal{N}(\bm{Z}; \bm{0},\bm{I})$;
			\STATE Calculate $\nabla_{\bm{H}_a^{i-1}}\log p\left(\bm{H}_a^{i-1}\vert\bm{Y}\right)$ via (\ref{equ.post_score_dis});
			\STATE $\epsilon_{i-1} = 2\sigma_{i-1}\left(r\|\bm{Z}\|_2/\|\nabla_{{\bm{H}}_a^{i-1}} \log p({\bm{H}}_a^{i-1}|\bm{Y})\|_2\right)^2$;
			\STATE $\bm{H}_a^{i-1} = {\bm{H}}_a^{i-1} + \epsilon_{i-1} \nabla_{{\bm{H}}_a^{i-1}} \log p({\bm{H}}_a^{i-1}|\bm{Y}) + \sqrt{2\epsilon_{i-1}} \bm{Z}$;
		\ENDFOR \quad \% \textit{Corrector: lines 7-12}
	\ENDFOR
	\ENSURE Estimated channel $\hat{\bm{H}}_a = \bm{H}_{a}^{0}$. 
\end{algorithmic} 
\end{algorithm}

\subsection{Data Detection}\label{S5.3}

Upon reviewing the formulas (\ref{equ.sys1}) and (\ref{equ.trans_jcedd}), the DD problem can be characterized as $\bm{Y}_d\! =\! \bm{X}_a\bm{H}_a\! +\! \bm{W}_d$. By transposing the formula and representing the model in a real-valued form, the signal transmission model can be reformulated as:
\begin{equation}\label{equ.DD_trans_model} 
	\mathring{\bm{Y}}_d = \mathring{\bm{H}}_a\mathring{\bm{X}}_a + \mathring{\bm{W}}_d,
\end{equation}
where $\mathring{\bm{Y}}_d$, $\mathring{\bm{H}}_a$, $\mathring{\bm{X}}_a$ and $\mathring{\bm{W}}_d$ are the transposed counterparts of the corresponding variables in real-valued form. Similar to the CE problem, assuming that the channel $\mathring{\bm{H}}_a$ is known, data symbols can be sampled from the posterior distribution $p(\mathring{\bm{X}}_a|\mathring{\bm{Y}}_d)$ by solving the SDE, which also involves calculating the prior score and the likelihood score of the data. Since the data symbols are uniformly sampled from a QAM constellation, the prior score $\nabla_{\mathring{\bm{X}}_a^{i}}\log p\big(\mathring{\bm{X}}_a^{i}\big)$ in the $i$-th iteration step can be calculated using Tweedie's formula \cite{Tweedie}, i.e., 
\begin{equation}\label{equ.prior_score_X} 
  \nabla_{\mathring{\bm{X}}_a^{i}} \log p(\mathring{\bm{X}}_a^{i}) = \frac{\mathbb{E}\big\{\mathring{\bm{X}}_a^{0}|\mathring{\bm{X}}_a^{i}\big\} - \mathring{\bm{X}}_a^{i}}{\tau_{i}^2},
\end{equation}
where $\tau_{i}\! =\! \tau\left(t\right)|_{t = \frac{i}{N}}$ is the scheduled perturbation noise for data symbol, and $\tau(t)\! =\! \tau_{\min}\big(\frac{\tau_{\max}}{\tau_{\min}}\big)^{t}$ is a predefined scheduling function. Additionally, the posterior distribution can be derived using Bayes' rule as: $p(\mathring{\bm{X}}_a^{0}|\mathring{\bm{X}}_a^{i})\! =\! \frac{p(\mathring{\bm{X}}_a^{i}|\mathring{\bm{X}}_a^{0})p(\mathring{\bm{X}}_a^{0})}{\mathop{\sum}_{\mathring{\bm{X}}_a^{0}\sim \mathcal{X}}p(\mathring{\bm{X}}_a^{i}|\mathring{\bm{X}}_a^{0})p(\mathring{\bm{X}}_a^{0})}$, where $p(\mathring{\bm{X}}_a^{0})\! =\! \frac{1}{\vert \mathcal{X} \vert_c}$. Then, the posterior mean is derived as
\begin{equation}\label{equ.post_mean_X} 
  \mathbb{E}\left\{\mathring{\bm{X}}_a^{0}|\mathring{\bm{X}}_a^{i}\right\} = \frac{\sum_{\mathring{\bm{X}}_a^{0}\sim \mathcal{X}} \mathring{\bm{X}}_a^{0}p(\mathring{\bm{X}}_a^{i}|\mathring{\bm{X}}_a^{0})}{\sum_{\mathring{\bm{X}}_a^{0}\sim \mathcal{X}} p(\mathring{\bm{X}}_a^{i}|\mathring{\bm{X}}_a^{0})}. 
\end{equation}
By substituting (\ref{equ.post_mean_X}) into (\ref{equ.prior_score_X}), the final prior score the data symbols at time step $i$ can be expressed as
\begin{equation}\label{eqFps} 
  \nabla_{\mathring{\bm{X}}_a^{i}}\! \log p(\mathring{\bm{X}}_a^{i})\! =\! \frac{1}{\tau_{i}^2}\!\! \left(\!\! \frac{\sum_{\mathring{\bm{X}}_a^{0}\sim \mathcal{X}} \mathring{\bm{X}}_a^{0} e^{-\frac{\|\mathring{\bm{X}}_a^{i} - \mathring{\bm{X}}_a^{0}\|^2}{2\tau_{i}^2}}}{\sum_{\mathring{\bm{X}}_a^{0}\sim \mathcal{X}} e^{-\frac{\|\mathring{\bm{X}}_a^{i}-\mathring{\bm{X}}_a^{0}\|^2}{2\tau_{i}^2}}}\! -\! \mathring{\bm{X}}_a^{i}\!\! \right)\!\! .\!
\end{equation}

\begin{table*}[!b]
\vspace*{-3mm}
\centering
\caption{Complexity of Algorithm~\ref{alg.jcedd}} 
\label{tab:complexity} 
\vspace*{-2mm}
\resizebox{0.8\textwidth}{!}{
\renewcommand{\arraystretch}{1.5}
\begin{tabular}{|c|c|l|c|c|}
	\hline
	\multicolumn{2}{|c|}{\textbf{Operation}} & \textbf{Lines} & \multicolumn{1}{c|}{\textbf{Complexity/Each Time}} & \multicolumn{1}{c|}{\textbf{Number of Times}} \\ \hline
	\multicolumn{2}{|c|}{Initialization} & Line 1 & $\mathcal{O}\left(K_aL_p^2M^3+L_pK_a^2M^3+\frac{2}{3}L_p^3M^3+2K_aM^2L_p^2\right)$ & 1 \\ \hline
	\multirow{2}{*}{SVD} & Channel & Lines 2 and 13 & $\mathcal{O}\left(2LK_a^2+K_a^3\right)$ & $1 + \left\lfloor {i^{*}}/{N_{\text{update}}} \right\rfloor$ \\ \cline{2-5}
	& Data & Line 7 & $\mathcal{O}\left(2K_aM^2+M^3\right)$ & $\left\lfloor {i^{*}}/{N_{\text{update}}} \right\rfloor$ \\	\hline	
	\multirow{2}{*}{\shortstack{Posterior Score \\ Calculation}} & Channel & Lines 4-5 & $\mathcal{O}(L^2M + K_a^2M + K_aF_{\text{net}})$ & $i^{*}(1+Q)$ \\ \cline{2-5}
	& Data & Lines 10-11 & $\mathcal{O}(M^2L_d + K_a^2L_d + 2K_aL_d\mathcal{\vert X \vert}_c)$ & $\left\lfloor {i^{*}}/{N_{\text{update}}} \right\rfloor(1+Q)N_X$ \\ \hline	
	\multicolumn{3}{|c|}{\textbf{Overall Complexity}} & \multicolumn{2}{c|}{Dominated by Lines 4-5 and Lines 10-11.} \\ \hline
\end{tabular}}
\vspace{-2mm}
\end{table*}

For the likelihood score, the equivalent likelihood function can be derived from the transmission model (\ref{equ.DD_trans_model}) as:
\begin{align}\label{eqELF} 
	\nabla_{\mathring{\bm{X}}_a^{i}}\log p\left(\mathring{\bm{Y}}_d\vert \mathring{\bm{X}}_a^{i}\right) =& \bm{V}_h\bm{\Sigma}_h^{\rm T} \left(\sigma_n^2\bm{I}_{2M} + \tau_i^2\bm{\Sigma}_h \bm{\Sigma}_h^{\rm T}\right)^{-1} \nonumber \\
	& \times \left(\bm{U}_h^{T}\mathring{\bm{Y}}_d-\bm{\Sigma}_h\bm{V}_h^{\rm T}\mathring{\bm{X}}_a^{i}\right),
\end{align}
where we have used the SVD $\mathring{\bm{H}}_a = \bm{U}_h\bm{\Sigma}_h\bm{V}_h^{\rm T}$ to reduce the complexity during the iteration. Finally, the posterior score of the data symbols can be calculated as 
\begin{align}\label{equ.post_score_x} 
  \nabla_{\mathring{\bm{X}}_a^{i}} \log p\left(\mathring{\bm{X}}_a^{i}\vert \mathring{\bm{Y}}_d\right) =&	\lambda_{x}\nabla_{\mathring{\bm{X}}_a^{i}} \log p\left(\mathring{\bm{Y}}_d\vert \mathring{\bm{X}}_a^{i}\right) \nonumber \\
	& + \nabla_{\mathring{\bm{X}}_a^{i}} \log p(\mathring{\bm{X}}_a^{i}) ,
\end{align}
where $\lambda_x\! =\! 2.5$ is an empirically chosen hyperparameter. 
The DD algorithm can also be derived based on PC-sampler. Due to space constraint, the detailed DD algorithm is omitted, but its basic iterative procedure is similar to that of Algorithm~\ref{alg:pc_ce}.

\begin{figure*}[!t]
\centering
\captionsetup{font={footnotesize}}
\includegraphics[width=0.8\textwidth]{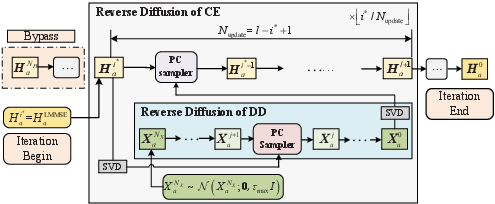}
\caption{Block diagram of the proposed PC sampler-based JCEDD algorithm.}
\label{fig:jcedd} 
\vspace{-2mm}	
\end{figure*}

\subsection{Overall Algorithm}\label{S5.4}

In the CE and DD algorithms above, the data $\bm{X}_a$ is assumed to be known during CE, while the channel $\bm{H}_a$ is assumed to be known during DD. In practical transmission, only the pilot $\bm{P}_a$ and the received signals $\bm{Y}$ are known. Therefore, we design an iterative algorithm to perform JCEDD.
The proposed PC sampler-based JCEDD method is depicted in Fig.~\ref{fig:jcedd}, and the algorithmic framework is summarized as Algorithm~\ref{alg.jcedd}. 

The issues we need to address are: 1)~A typical SDE sampling algorithm starts by sampling from random noise, resulting in a high number of iterations. 2)~In the existing ALD-based JCEDD \cite{nicolas-24icassp}, each CE iteration is followed by one DD iteration. Due to the different distributions of the channel and data symbols, their convergence speeds are inconsistent, causing the DD to fall into local optima before the completion of the CE iterations. We address these issues by the following improvements in the proposed sampling mechanism.
\begin{itemize}
\item We conduct the LMMSE estimation \cite{oamp-20tsp} of the channel $\bm{H}_a$ using $\bm{Y}_p\! =\! \bm{P}_a\bm{H}_a\! +\! \bm{W}_p$ and determine the perturbation noise variance that matches the estimated channel variance as the starting point for SDE sampling. Let the estimated covariance matrix of the channel $\bm{h}_a\! =\! \text{vec}\big(\bm{H}_a^{\rm T}\big)\! \in\! \mathbb{C}^{K_a M}$ be $\bm{R}_\Delta$. Then the starting point $i^{*}$ of the SDE iteration is obtained by identifying $\sigma_{i}$ that is the closest to the average channel variance according to: 
\begin{equation}\label{equ.init_i} 
  i^{*} = \arg \min\limits_{\forall i} \bigg\vert \sigma_{i} - \frac{1}{K_aM} \sum_{l=1}^{K_a M} \left[\bm{R}_{\Delta}\right]_{l,l}\bigg\vert .
\end{equation}
By reshaping the estimated $\hat{\bm{h}}_a$ into matrix $\hat{\bm{H}}_a$, the initial DD result can be obtained by applying zero forcing (ZF) to (\ref{equ.DD_trans_model}), i.e., $\hat{\bm{X}}_a\! =\! \bm{Y}_d\bm{H}_a^{\rm H}\big(\bm{H}_a\bm{H}_a^{\rm H}\big)^{-1}$.

\item We design an asynchronous alternating iteration mechanism. For channel datasets with complex distributions, the calculation of posterior score is more complicated than DD that relies solely on model-based iterations with closed-form expression. 
We update the DD results after every $N_{\text{update}}$ channel iterations (Line 6 in Algorithm~\ref{alg.jcedd}). To ensure the optimal DD results under the current channel error, we conduct $N_X$ steps of PC sampler each time the data symbols are updated (Lines 9-12). 
\end{itemize}

\begin{algorithm}[t]
\caption{PC Sampler-Based JCEDD Algorithm}
\label{alg.jcedd} 
\begin{algorithmic}[1] 
	\REQUIRE ${\bm{S}_a}$, ${\bm{Y}}$, $\sigma_n^2$, pre-trained denoising network $\bm{s}_{\bm{\theta}^{*}}$, SDE noise scheduling functions $\sigma(t)$ and $\tau(t)$, step numbers $N_H$ and $N_D$, hyper-parameters $\lambda_{h}$ and $\lambda_{x}$; 
	\STATE \textbf{Initialize:} Acquire initial CE $\hat{\bm{H}}_a$ via LMMSE, initial DD $\hat{\bm{X}}_a$ via ZF, and start point $i^{*}$ via (\ref{equ.init_i});
	\STATE \textbf{Calculate:} $\bm{S}_a = \big[\bm{P}_a^{\rm T}, \hat{\bm{X}}_a^{\rm T}\big]^{\rm T} = \bm{U}\bm{\Sigma}\bm{V}^{\rm T}$;
	\FOR{$i=i^{*}$ to $1$}
		\STATE Calculate $\nabla_{\bm{H}_a^{i}}\log p\left(\bm{H}_a^{i}\vert\bm{Y}\right)$ via (\ref{equ.post_score_dis});
		\STATE $\bm{H}_a^{i-1} = \text{PC-sampler}\left(i, \bm{H}_a^{i}, \nabla_{\bm{H}_a^{i}}\log p\left(\bm{H}_a^{i}\vert\bm{Y}\right) \right)$;
		\IF{$i \% N_{\text{update}} == 0$}
			\STATE {Calculate} $\bm{H}_a^{i-1} = \bm{U}_{h}\bm{\Sigma}_{h}\bm{V}_{h}^{\rm T}$;
			\STATE {Initialize} $\mathring{\bm{X}}_a^{N_X}\sim \mathcal{N}\left(\mathring{\bm{X}}_a^{N_X};\bm{0},\tau_{\max}^2\bm{I}\right)$;
			\FOR{$j=N_X$ to $1$}
				\STATE Calculate $\nabla_{\mathring{\bm{X}}_a^{j}}\log p\left(\mathring{\bm{X}}_a^{j}\vert \mathring{\bm{Y}}_d\right)$ via (\ref{equ.post_score_x});
				\STATE $\bm{X}_a^{j-1} =\text{PC-sampler}\left(j, \bm{X}_a^{j}, \nabla_{\mathring{\bm{X}}_a^{j}}\log p\left(\mathring{\bm{X}}_a^{j}\vert \mathring{\bm{Y}}_d\right)\right)$;
			\ENDFOR
			\STATE {Calculate} $\bm{S}_a = \big[\bm{P}_a^{\rm T}, \left(\bm{X}_a^0\right)^{\rm T}\big]^{\rm T} = \bm{U}\bm{\Sigma}\bm{V}^{\rm T}$;
		\ENDIF
	\ENDFOR
	\ENSURE Estimated channel and data $\hat{\bm{H}}_a\! =\! \bm{H}_{a}^{0}$ and $\hat{\bm{X}}_a\! =\! \bm{X}_{a}^{0}$.
\end{algorithmic} 
\end{algorithm}

\subsection{Computational Complexity Analysis}\label{S5.5}

The complexity analysis of Algorithm \ref{alg.jcedd} is detailed in Tab. \ref{tab:complexity}. The channel is initialized using the LMMSE method before the iterative process begins. For the channel PC-sampler, $i^{\star}$ iterations are necessary. Each PC iteration consists of one predictor step and $Q$ corrector steps, resulting in a total of $i^{\star}(1\! +\! Q)$ computations of posterior scores, as described in Lines 4-5 of Algorithm \ref{alg.jcedd}. Furthermore, data is updated after every $N_{\text{update}}$ channel iterations, as outlined in Lines 6-14.

During each data update, $N_X$ data iterations are conducted using another PC sampler. Each iteration includes one predictor step and $Q$ corrector steps, leading to a total of $\big\lfloor \frac{i^{\star}}{N_{\text{update}}} \big\rfloor (1\! +\! Q) N_X$ computations of posterior scores for data.
Additionally, SVD is utilized to reduce computational complexity. Specifically, $1\! +\! \big\lfloor \frac{i^{\star}}{N_{\text{update}}} \big\rfloor$ SVD operations are performed for the channel in Lines 2 and 13, and $\big\lfloor \frac{i^{\star}}{N_{\text{update}}} \big\rfloor$ SVD operations are needed for the data in Line 7.

Specifically, computational complexity of calculating each channel likelihood score function is $\mathcal{O}(L^2M\! +\! K_a^2M)$, while the complexity of the channel prior score function is $\mathcal{O}(K_aF_{\text{net}})$, where $F_{\text{net}}$ represents the computational cost of one time inference of the score network $\boldsymbol{s}_{\boldsymbol{\theta}}(\boldsymbol{h}(t),t)$. The data likelihood score's complexity is $\mathcal{O}(M^2L_d\! +\! K_a^2L_d)$, and the data prior score's complexity is $\mathcal{O}(2K_aL_d\vert \mathcal{X} \vert_c)$.

For typical simulation parameters, e.g., $N_{\text{update}}\! =\! 50$, $\big\lfloor \frac{i^{\star}}{N_{\text{update}}} \big\rfloor\! =\! 5\! \sim\! 15$, $Q\! =\! 3$, $N_X\! = \!1500$ and $N_H\! =\! 1500$, the overall complexity of Algorithm 2 is primarily determined by the score function computations in lines 4-5 and 10-11. Specifically, the neural network NCSN++ \cite{song-21sde}, employed to calculate the channel's prior score function, markedly impacts the inference complexity. For complex channel distributions, utilizing a moderately complex network can balance performance with computational demands, providing an avenue for algorithm optimization. Additionally, given that the proposed approach includes multiple PC-sampler iterations, the use of higher-order SDE or ordinary differential equation (ODE) solvers \cite{ddim-21iclr, dpmv3_23nips} is essential for further reducing complexity in engineering applications.

\section{Numerical Results}\label{S6}
In this section, we first detail the experimental setup, including simulation parameters, network configurations, channel datasets, training conditions, and comparative algorithms. We then assess the performance of the Transformer-based AUD as well as the generative diffusion models driven JCEDD.

\subsection{Experimental Setups}\label{S6.1}

We consider a typical uplink random access scenario with a total of $K\! =\! 128$ potential users\footnote{Future networks are expected to support millions of devices per square kilometer. However, such high user density can only be achieved within specific time-frequency resources. For anticipated 6G massive communications, it is generally necessary to divide limited resources into different time-frequency resource groups. For example, within the coverage area of a single macro base station (approximately $S = \pi r^2 = \pi \times 0.5^2 = 0.7854 \, \text{km}^2$), 128 users can be grouped to share the same resource block, such as 12 subcarriers and 14 OFDM symbols. Considering a system with 1 frame = 10 ms = 140 OFDM symbols and 512 resource blocks (providing approximately 92.16 MHz bandwidth for 15 kHz subcarrier spacing), the system can theoretically support up to $512 \times 10 \times 128 = 655,360$ UEs, corresponding to a user density of $\frac{655,360 \, \text{UEs/sector} \times 3 \, \text{sectors}}{0.7854 \, \text{km}^2} \approx 2.5 \times 10^6 \, \text{UEs/km}^2$ for the $100 \, \text{MHz} \times 10 \, \text{ms}$ time-frequency resources, effectively meeting the massive communication requirement. }, where each user's activity follows a Bernoulli distribution with an active probability of $0.1$. The BS employs a UPA with $M_x\! =\! M_y\! =\! 8$ antennas on each dimension, while each user operates with a single antenna. Each element of the pilot sequence $\tilde{\bm{p}}_k$, $\forall k$, and the data symbols $\bm{x}_k$ are randomly sampled from 4QAM constellations. The maximum pilot sequence length is $L_p^{\text{max}}\! =\! 28$. We consider a 3GPP-compliant channel model \cite{38.901} generated using Quadriga \cite{QuaDRiGa}, simulating the 3GPP 38.901 UMa NLOS scenario at a carrier frequency of 3.5 GHz. The minimum pilot sequence length is $L_p^{\text{min}}\! =\! 8$ for Rayleigh channel, and $L_p^{\min}\! =\! 10$ for 3GPP channel, since 3GPP channel is more complicated and it needs more pilots to estimate AUS.

\subsubsection{VPL-AUDNet Setup}
A total of 64,000 channel samples are generated using Quadriga. For each training batch, the pilot lengths for the UEs are randomly selected from $[L_p^{\text{min}},\, L_p^{\text{max}}]$. In addition to the input preprocessing layer and output layer, the VPL-AUDNet contains 5 heterogeneous Transformer encoder layers, each featuring a hidden state sequence length of $d\! =\! 512$ and 8 attention heads. Training is conducted over 200 epochs, with the learning rate warming up to $10^{-4}$ during the initial 15 epochs and then decaying to 0.1 of this original value every 75 epochs. The batch size is set to 128.

\subsubsection{Channel Prior Score Network Setup}
The NSCN++ architecture \cite{song-21sde} is employed to model the prior score function $\bm{s}_{\bm{\theta}}(\bm{h}(t),t)$. A 3GPP training dataset consisting of 160,000 channel samples is utilized. The time step $t$ is uniformly sampled from the interval $[0,\, 1]$, and the perturbation noise variance is determined using $\sigma(t)\! =\! \sigma_{\min}\left(\frac{\sigma_{\max}}{\sigma_{\min}}\right)^{t}$, with $\sigma_{\text{max}}\! =\! 30$ and $\sigma_{\text{min}}\! =\! 0.01$. The network is trained with a learning rate of $10^{-5}$ and a batch size of 64, over 20 epochs.

During the testing phase, the reverse SDE process for CE is discretized into $N_H\! =\! 1,500$ steps. Given that an LMMSE initialization is employed, only $i^{*}$ of the $N_H$ steps are executed. The reverse SDE process for DD is also discretized into $N_X\! =\! 1,500$ steps, using a noise scheduling function defined as $\tau(t)\! =\! \tau_{\min}\big(\frac{\tau_{\max}}{\tau_{\min}}\big)^{t}$, where $\tau_{\text{max}}\! =\! 1$ and $\tau_{\text{min}}\! =\! 0.01$. Additionally, the hyperparameters are set to $\lambda_h\! =\! \lambda_x\! =\! 2.5$.

\subsubsection{Evaluation Metrics}
We utilize activity error probability (AEP), normalized mean square error (NMSE) and bit error rate (BER) to evaluate the performance of AUD, CE and DD, respectively. 
These metrics are defined respectively as $\text{AEP}\! =\! \mathbb{E}\big\{\frac{1}{K}\sum_{k=1}^{K} \vert \alpha_k\! -\! \hat{\alpha}_{k}\vert\big\}$, $\text{NMSE}\! =\! \mathbb{E}\big\{\frac{\| \bm{H}_a\! -\! \hat{\bm{H}}_a\|^2}{\| \bm{H}_a\|^2}\big\}$ and $\text{BER}\! =\! \mathbb{E}\big\{\frac{E_b}{K_a L_d \log_2\vert\mathcal{X}\vert_c}\big\}$, where $E_b$ denotes the total number of error bits for active UEs.

\subsubsection{Comparative Schemes}
For AUD, the proposed method is compared with two baseline algorithms: AMP \cite{kml-tsp} and OMP \cite{ql-tvt}. Additionally, we also assess the performance of the VPL-AUDNet scheme trained with a fixed pilot length. For CE and DD, the following algorithms are included in the comparisons:

\noindent
$\bullet$~\textbf{LS-CE \& Perfect-Data:} The data $\bm{X}_a$ is perfectly known, and jointly utilize pilots $\bm{P}_a$ and data $\bm{X}_a$ for LS CE;

\noindent
$\bullet$~\textbf{Perfect-CSI \& ZF-DD:} The CSI $\bm{H}_a$ is perfectly known, and employ ZF for DD;

\noindent
$\bullet$~\textbf{Perfect-CSI \& OAMP-DD:} The CSI $\bm{H}_a$ is perfectly known, and utilize OAMP \cite{oamp-20tsp} for DD;

\noindent
$\bullet$~\textbf{Pilot LS-CE \& ZF-DD:} Utilize pilots to perform LS CE, and use the estimated channel for ZF DD;

\noindent
$\bullet$~\textbf{Pilot LS-CE \& OAMP-DD:} Utilize pilots to perform LS CE, and use the estimated channel for OAMP \cite{oamp-20tsp} DD;

\noindent
$\bullet$~\textbf{Iter LMMSE-CE \& OAMP-DD:} Employ LMMSE for CE and OAMP for DD, with CE and DD iteratively executed to enhance the performance of each other \cite{oamp-20tsp};

\noindent
$\bullet$~\textbf{Iter Langevin CE \& DD:} Conduct iterative CE and DD using ALD \cite{nicolas-24icassp};

\noindent
$\bullet$~\textbf{SDE-CE \& Perfect-Data:} The data $\bm{X}_a$ is perfectly known, and apply SDE solver (i.e., PC sampler) for CE, which corresponds to Algorithm~\ref{alg:pc_ce};

\noindent
$\bullet$~\textbf{Perfect-CSI \& SDE-DD:} The CSI $\bm{H}_a$ is known, and apply SDE solver for DD;

\noindent
$\bullet$~\textbf{Our Iter SDE CE \& DD:} The proposed Algorithm~\ref{alg.jcedd}.

\begin{figure}[!t]
\centering
\captionsetup{font={footnotesize}}
	\hspace{-4mm}
	\subfigure[Rayleigh channel]{
		\includegraphics[width=.25\textwidth,keepaspectratio]{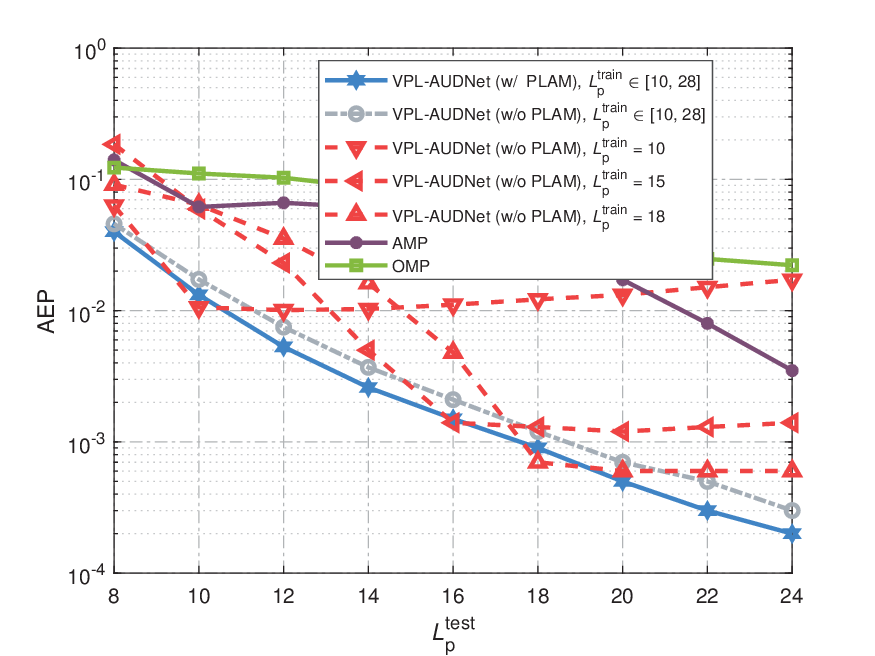}
		\label{fig_aud_1}
	}
	\hspace{-8mm}
	\subfigure[3GPP channel]{
		\includegraphics[width=.25\textwidth,keepaspectratio]{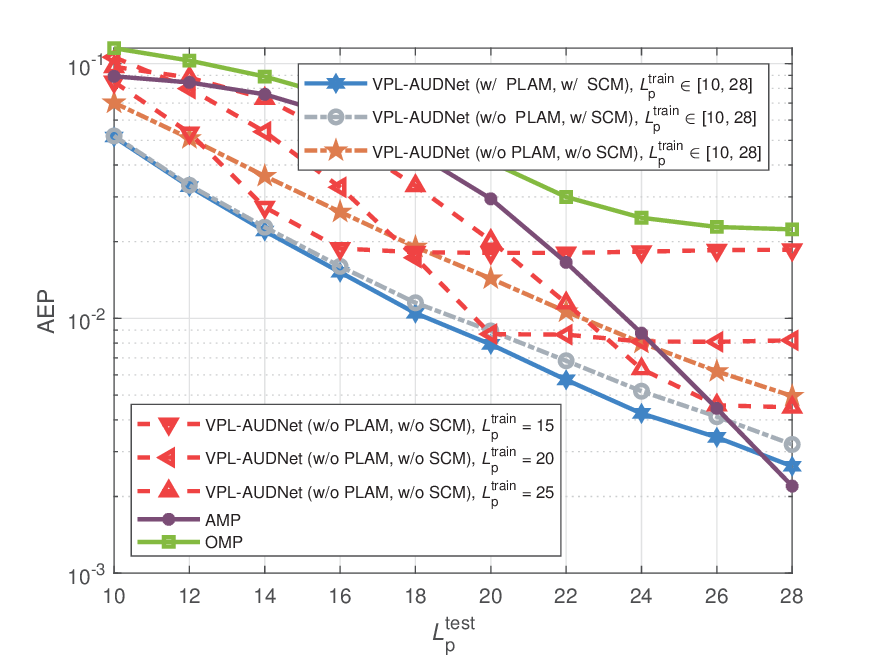}
		\label{fig_aud_2}
	}
	\vspace*{-3mm}
\caption{AEP performance with respect to the transmission pilot length of different AUD schemes.}
\label{fig_aud_snr} 
\vspace*{-5mm}
\end{figure}

\subsection{Performance of AUD}\label{S6.2}

Fig.~\ref{fig_aud_1} illustrates the AUD performance of various schemes as the functions of the transmission pilot length, $L_p^{\text{test}}$, under i.i.d. Rayleigh channels, where the channel between each potential UE and the BS is modeled as $\bm{h}_k\! \sim\! \mathcal{N}_c(\bm{h}_k;\bm{0},\bm{I}_M), \forall k$. The simulation is conducted under the received $\text{SNR}\! =\! \frac{\mathbb{E}\left\{\|\bm{S}\bm{H}\|^2\right\}}{\mathbb{E}\left\{\|\bm{W}\|^2\right\}}\! =\! 10$\,dB. The results indicate that that the VPL-AUDNet-based schemes outperform traditional CS algorithms based on AMP and OMP, in terms of AEP. Furthermore, the VPL-AUDNet models trained with fixed pilot lengths and without the PLAM module, e.g., VPL-AUDNet (w/o PLAM), $L_p^{\text{train}}\! =\! 15$, exhibit degraded performance when tested on pilot lengths different from the training pilot length\footnote{For models trained with a pilot length of $L_p^{\text{train}}$, if the test length is shorter than the training length, the received signals are padded with zeros to align with $L_p^{\text{train}}$ before being input into the VPL-AUDNet network. Conversely, if the test length exceeds the training length, the received signals are truncated to $L_p^{\text{train}}$ before network input.}. Conversely, VPL-AUDNet (w/o PLAM), $L_p^{\text{train}}\! \in\! [10,\, 28]$ exhibit better generalization performance. Moreover, the proposed VPL-AUDNet (w/ PLAM), $L_{\text{train}}\! \in\! [8,\, 24]$, outperforms VPL-AUDNet (w/o PLAM), $L_{\text{train}}\! \in\! [8,\, 24]$, demonstrating the effectiveness of the PLAM. Additionally, under i.i.d. Rayleigh channels, the proposed VPL-AUDNet scheme achieves an AEP performance close to $10^{-2}$ even when the pilot number is less than the average number of UEs ($L_p^{\text{test}}\! \le\! 12$), whereas the AMP and OMP algorithms exhibit a AEP of $10^{-1}$. These results demonstrate that the proposed scheme achieves optimal overall performance across various pilot lengths within a single model, thereby confirming its effectiveness.

Fig.~\ref{fig_aud_2} compares the AUD performance of various schemes as  the functions of the number of transmission pilots $L_p^{\text{test}}$ under 3GPP channels, given a received SNR of 10\,dB. The results show that the performance curves of the different schemes are similar to the trends observed under Rayleigh channels. Compared to traditional CS algorithms like AMP and OMP, the VPL-AUDNet-based scheme achieves better performance with significantly reduced pilot length by avoiding completely CE. Additionally, a comparison between the curves of VPL-AUDNet (w/o PLAM, w/ SCM), $L_p^{\text{train}}\! =\! [10,\, 28]$, and VPL-AUDNet (w/o PLAM, w/o SCM), $L_p^{\text{train}}\! =\! [10,\, 28]$, reveals that the SCM significantly enhances AEP performance under 3GPP channels. Moreover, VPL-AUDNet (w/ PLAM, w/ SCM), $L_p^{\text{train}}\! =\! [10,\, 28]$ scheme attains optimal performance at low pilot numbers. 

\begin{figure}[!h]
\vspace{-5mm}
\centering
\captionsetup{font={footnotesize}}
\includegraphics[width=0.89\columnwidth]{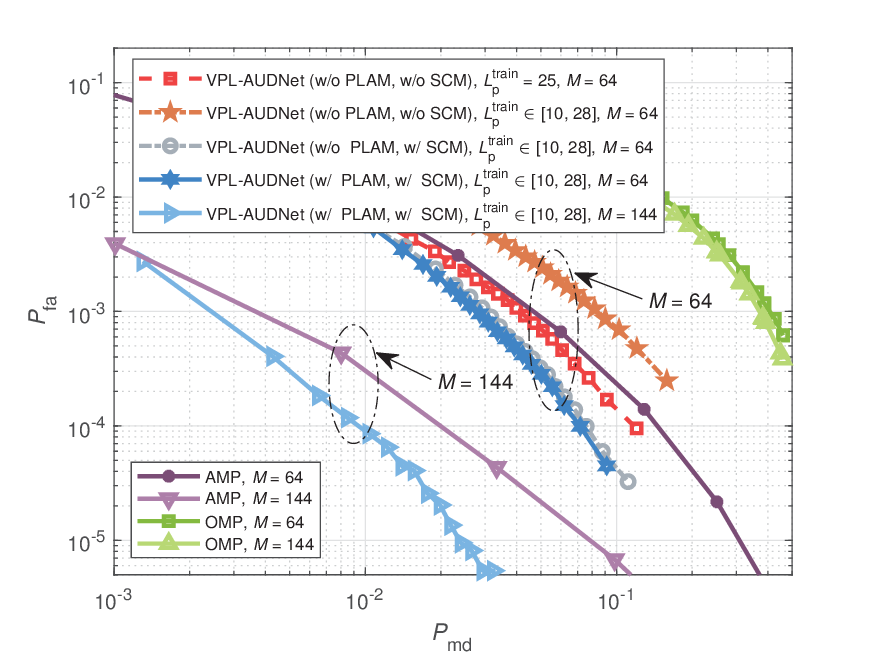}
\vspace{-2mm}
\caption{$P_{\text{fa}}\text{-}P_{\text{md}}$ performance of different AUD schemes under 3GPP channel.}
\label{fig_aud_3} 
\vspace{-1mm}
\end{figure}

Fig.~\ref{fig_aud_3} depicts the curves for the probability of false alarm ($P_{\text{fa}}$) and the probability of missed detection ($P_{\text{md}}$) for various schemes under 3GPP channels. Simulations are conducted with a received SNR of 10\,dB and a test pilot length of $L_p^{\text{test}}\! =\! 25$. All VPL-AUDNet models are 
trained with an antenna number of $M\! =\! 64$. The results indicate that the proposed VPL-AUDNet (w/ PLAM, w/ SCM), $L_p^{\text{train}}\! =\! [10,\, 28]$, achieves an optimal trade-off between $P_{\text{fa}}$ and $P_{\text{md}}$. Moreover, testing the model trained with $M\! =\! 64$ antennas on the channels with $M\! =\! 144$ antennas results in further improvements in $P_{\text{fa}}\text{-}P_{\text{md}}$ performance, demonstrating the scheme's robust generalization capability across varying antenna configurations. In contrast, the performance gains for OMP related to an increased number of antennas are less significant.

\begin{figure}[!h]
	\vspace*{-5mm}
	\captionsetup{font={footnotesize}}
	\begin{center}
		\includegraphics[width = 0.9\columnwidth]{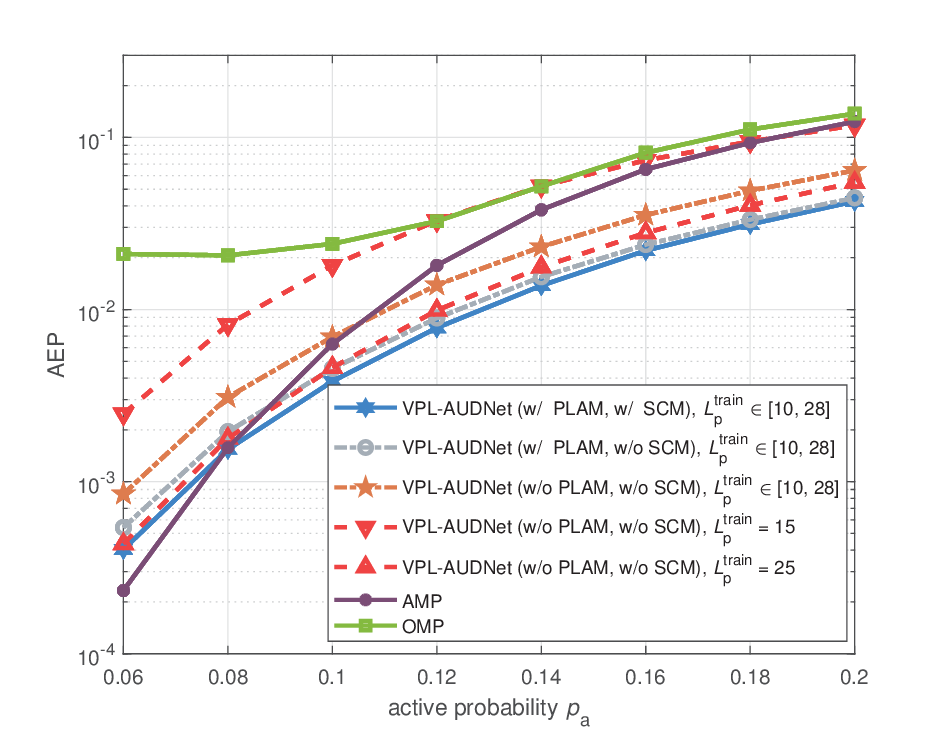}
	\end{center}
	\vspace*{-4mm}
	\caption{AEP performance with respect to the active probability of different AUD schemes under the 3GPP channel.}
	\label{fig.rl_aud_vary_ka} 
	\vspace*{-2mm}
\end{figure}

{Fig. \ref{fig.rl_aud_vary_ka} compares the AEP performance curves of different schemes as a function of active probability of UEs $p_a$ under the 3GPP channel. The simulation assumes a received SNR of 10 dB, a total potential number of UEs of 128, a user activation probability of $[0.06,  0.2]$, and a test pilot length of $ L_p^{\text{test}} = 25 $. The simulation results indicate that although AMP achieves marginally better AEP performance when the active probability is low, its performance deteriorates more rapidly as the active probability increases. In contrast, our scheme based on VPL-AUDNet exhibits a more gradual performance degradation, indicating better generalization capabilities in adapting to variations in activity probability.}

\begin{figure}[!h]
	\vspace{-3mm}
	\begin{center}
		\includegraphics[width = 0.9\columnwidth]{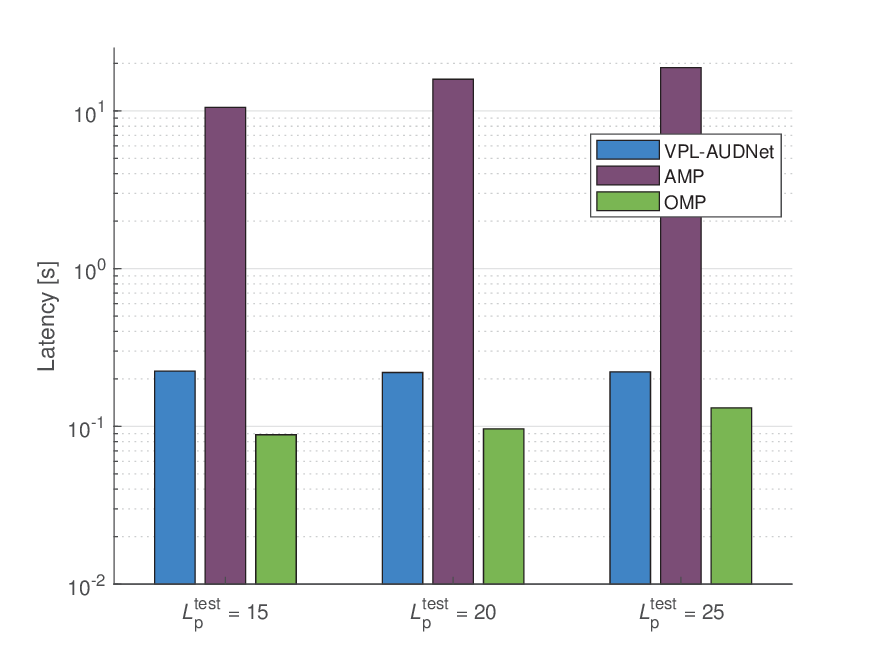}
	\end{center}
	\vspace*{-5mm}
	\captionsetup{font={footnotesize}, labelsep=period} 
	\caption{Computation latency of AUD schemes under different pilot lengths.}
	\label{rl_fig_aud_latency} 
	\vspace*{-2mm}
\end{figure}

Fig. \ref{rl_fig_aud_latency} illustrates the computational latency required to process a batch of 128 samples across different schemes with varying pilot lengths. The results show that the computational latency of the VPL-AUDNet scheme remains approximately constant regardless of the pilot length due to a specialized receiver structure accommodating variable pilot length inputs. In contrast, as the pilot length increases, compressed sensing algorithms based on AMP or OMP experience increased computational complexity because of the corresponding growth in the dimension of the sensing matrix. The overall computational delay of the proposed VPL-AUDNet scheme is lower than that of the AMP scheme but higher than that of the OMP scheme. Nevertheless, the AUD performance of the VPL-AUDNet scheme is generally superior to that of OMP, highlighting the advantages of a neural network-based approach.
\subsection{Performance of JCEDD}\label{S6.3}

\begin{figure}[!h]
	\centering
	\hspace{-6mm}
	\subfigure[]{\label{fig.rayleigh_nmse}%
		\includegraphics[width=.25\textwidth]{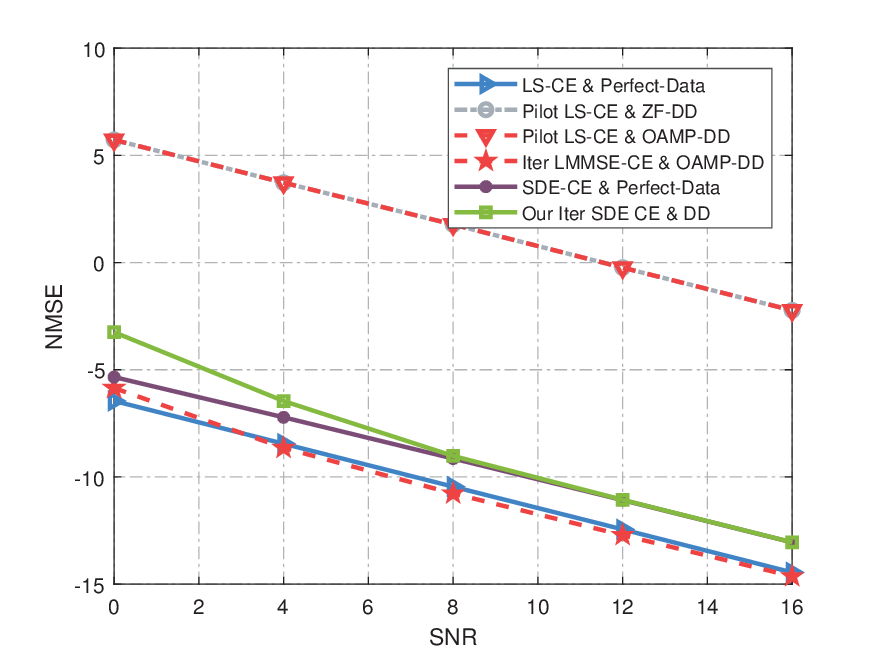}}
	\hspace{-6mm}
	\subfigure[]{\label{fig.rayleigh_ber}%
		\includegraphics[width=.25\textwidth]{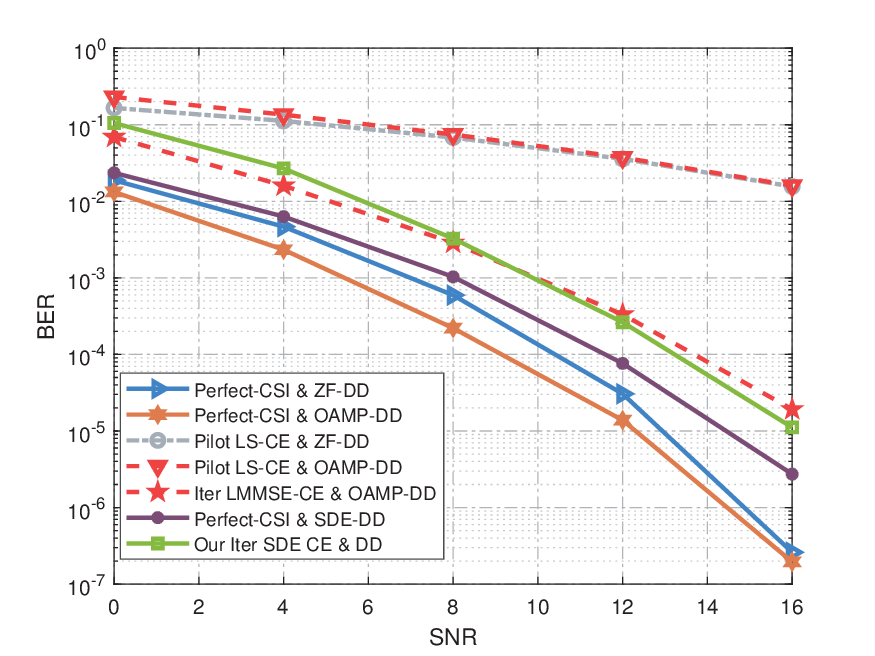}}
	\captionsetup{font={footnotesize}}
	\vspace{-2mm}
	\caption{NMSE and BER performance with respect to the received SNR for different CE and DD algorithms under the MIMO Rayleigh channel.}
	\label{fig.rayleigh_vs_snr}
\end{figure}

In this subsection's simulations, unless otherwise specified, the BS is equipped with $M\! =\! 64$ antennas, and a total of $K_a\! =\! 12$ active UEs access the system simultaneously. Moreover, apart from Fig. \ref{fig.rayleigh_vs_snr}, which utilizes a Rayleigh channel model, all other simulations are conducted using the 3GPP channel model.

\begin{figure}[!h]
\centering
	\captionsetup{font={footnotesize}}
	\hspace{-6mm}
	\subfigure[NMSE]{
		\includegraphics[width=.25\textwidth,keepaspectratio]{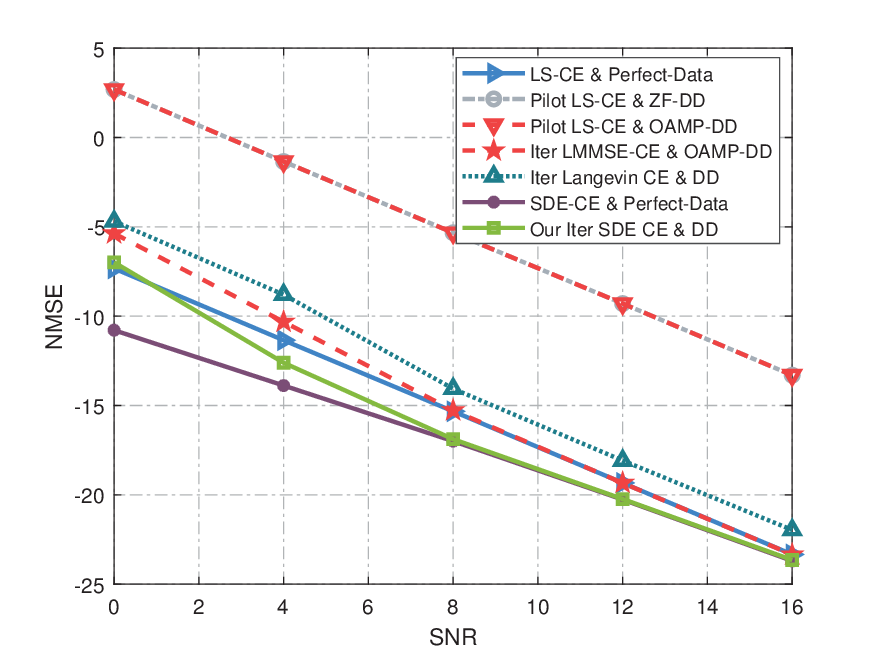}
		\label{fig_jcedd_1a}
	}
	\hspace{-6mm}
	\subfigure[BER]{
		\includegraphics[width=.25\textwidth,keepaspectratio]{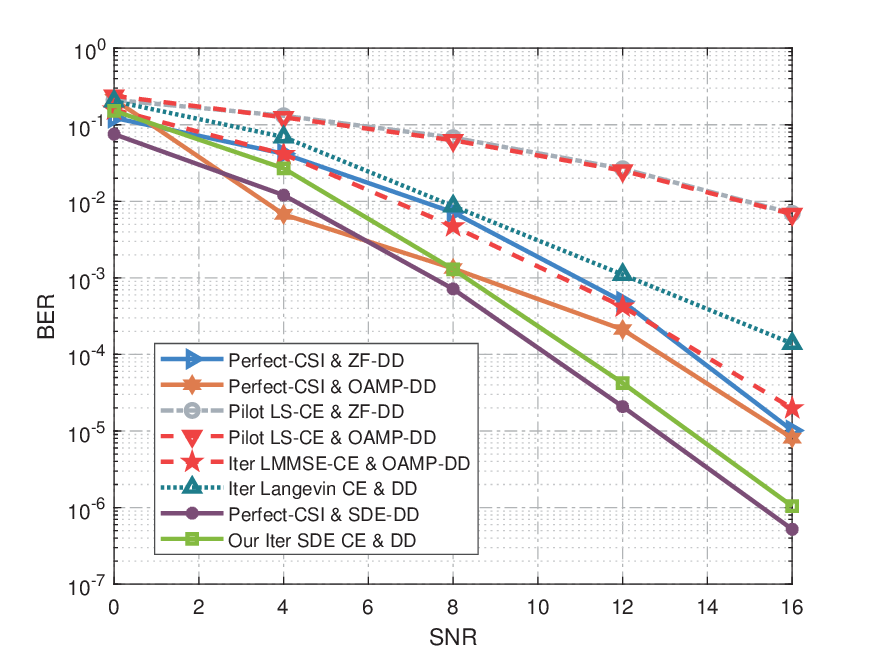}
		\label{fig_jcedd_1b}
	}
\vspace*{-2mm}
\caption{NMSE and BER performance with respect to the received SNR for different CE and DD algorithms under 3GPP channels.}
\label{fig_jcedd_1} 
\vspace*{-2mm}
\end{figure}

Fig.~\ref{fig_jcedd_1} illustrates the performance curves of CE and DD under varying SNRs for different schemes. In the simulations, each user is allocated a pilot length of $L_p\! =\! 15$ and a data length of $L_d\! =\! 50$. As shown in Fig.~\ref{fig_jcedd_1a}, the iterative CE and DD schemes (i.e., \textbf{Iter Langevin CE \& DD}, \textbf{Iter LMMSE-CE \& OAMP-DD}, and \textbf{Our Iter SDE CE \& DD}) achieve an NMSE gain of approximately 7 and 10\,dB when compared to the non-iterative \textbf{Pilot LS-CE \& ZF-DD}, and \textbf{Pilot LS-CE \& OAMP-DD} schemes. This indicates a significant gain in CE performance by using detected data to refine CE. As SNR increases, \textbf{Iter LMMSE-CE \& OAMP-DD}'s performance approaches that of \textbf{LS-CE \& Perfect-Data}. Similarly, the performance of \textbf{Our Iter SDE CE \& DD} (Algorithm \ref{alg.jcedd}) matches \textbf{SDE-CE \& Perfect-Data} (Algorithm \ref{alg:pc_ce}) for $\text{SNR}\! \ge\! 8$\,dB. When the data is not perfectly known, \textbf{Our Iter SDE CE \& DD} outperforms \textbf{Iter LMMSE-CE \& OAMP-DD} because it more effectively utilizes the prior distribution of the channel through a learned prior score function, whereas \textbf{Iter LMMSE-CE \& OAMP-DD} relies only on the channel's second-order statistical information via LMMSE. Additionally, \textbf{Our Iter SDE CE \& DD} implements a PC sampler to enhance SDE solution performance beyond what \textbf{Iter Langevin CE \& DD} algorithm can achieve with only a corrector. 

Fig.~\ref{fig_jcedd_1b} compares the BER performance of various algorithms. Similar to Fig.~\ref{fig_jcedd_1a}, the non-iterative \textbf{Pilot LS-CE \& ZF-DD} and \textbf{Pilot LS-CE \& OAMP-DD} algorithms perform worse than the iterative algorithms. \textbf{Our Iter SDE CE \& DD} scheme achieves performance close to that of \textbf{Perfect-CSI \& SDE-DD} and surpasses the \textbf{Iter LMMSE-CE \& OAMP-DD} and \textbf{Iter Langevin CE \& DD} algorithms. At a BER of $10^{-4}$, \textbf{Our Iter SDE CE \& DD} offers a 3\,dB SNR gain over \textbf{Iter LMMSE-CE \& OAMP-DD}. Furthermore, \textbf{Perfect-CSI \& SDE-DD} outperforms \textbf{Perfect-CSI \& OAMP-DD} and \textbf{Perfect-CSI \& ZF-DD} algorithms. This is likely because the channel matrix elements for the 3GPP scenario do not meet the i.i.d. distribution assumptions for DD tasks, causing OAMP algorithms to underperform. On the other hand, the ZF algorithm does not exploit the prior distribution information of the constellation.  In contrast, \textbf{Perfect-CSI \& SDE-DD} thoroughly leverages the prior distribution of the data and the likelihood information of the received signals, resulting in superior performance.

In Fig. \ref{fig.rayleigh_vs_snr}, we conducted simulations using the MIMO Rayleigh channel model, where the channel prior is assumed to follow $\mathcal{N}(\bm{0}, \bm{I})$. Instead of utilizing a neural network-learned score, we substituted the derived prior score (\ref{equ.prior_score_rayleigh}) into the process of Algorithm 2. The results indicate that \textbf{Our Iter SDE CE \& DD} scheme outperforms traditional two-stage algorithms such as \textbf{Pilot LS-CE \& ZF-DD} and \textbf{Pilot LS-CE \& OAMP-DD}. However, it is less effective compared to the \textbf{Iter LMMSE-CE \& OAMP-DD} algorithms. This is because, in a simple Rayleigh channel distribution, the mean and covariance matrix—representing first and second-order statistical information—are adequate to capture the channel's distribution characteristics. Consequently, the LMMSE-CE and OAMP-DD achieve near-optimal performance, while the prior score offers limited additional information. Additionally, numerical errors in solving the SDE process can affect the scheme's final performance. This suggests that the proposed scheme excels with more complex channel distributions, whereas existing methods already perform well in simple Rayleigh channels.

\begin{figure}[!t]
\centering
	\captionsetup{font={footnotesize}}
	\hspace{-6mm}
	\subfigure[NMSE]{
		\includegraphics[width=.25\textwidth,keepaspectratio]{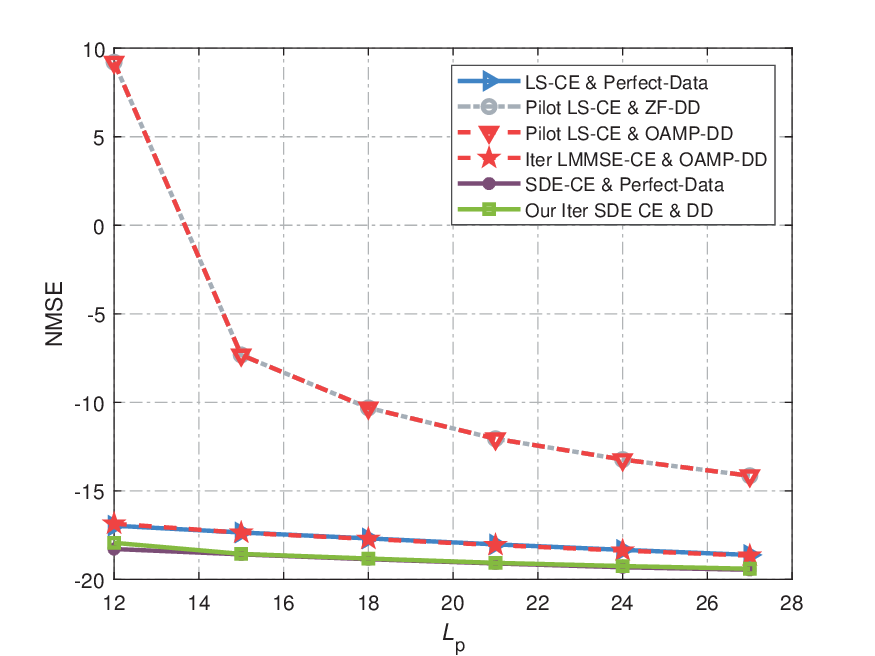}
		\label{fig_jcedd_2a}
	}
	\hspace{-6mm}
	\subfigure[BER]{
		\includegraphics[width=.25\textwidth,keepaspectratio]{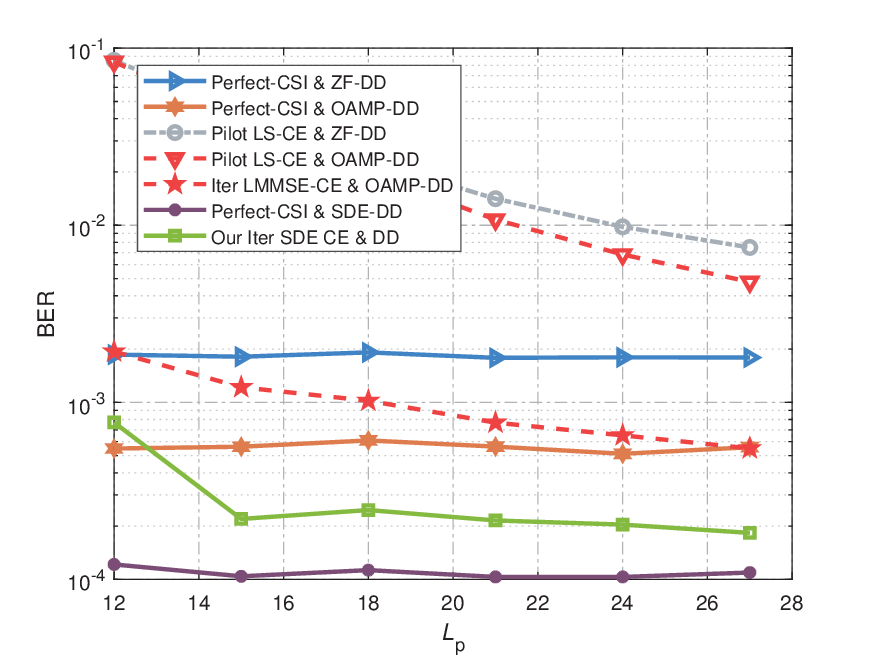}
		\label{fig_jcedd_2b}
	}
\vspace*{-2mm}
\caption{NMSE and BER performance with respect to the pilot symbol length $L_p$ for different CE and DD algorithms.}
\label{fig_jcedd_2} 
\vspace*{-5mm}
\end{figure}

Fig.~\ref{fig_jcedd_2} compares the performance curves of CE and DD as the functions of pilot length $L_p$ for different schemes. In the simulation, each user's data length is $L_d\! =\! 50$, and the received SNR at the BS is set to 10\,dB. As shown in Fig.~\ref{fig_jcedd_2a}, when $L_p\! >\! 12$, \textbf{Our Iter SDE CE \& DD} can achieve the performance of \textbf{SDE-CE \& Perfect-Data}, whereas the LS algorithm, which relies solely on pilots, exhibits a notable performance gap relative to iterative CE and DD algorithms. This suggests that additional computation can potentially reduce communication pilot overhead. Also \textbf{Iter LMMSE-CE \& OAMP-DD} and \textbf{LS-CE \& Perfect-Data} exhibit similar performance, indicating that the estimated data is sufficiently robust to enhance CE. Fig.~\ref{fig_jcedd_2b} shows that the BER performance of \textbf{Our Iter SDE CE \& DD} is considerably better than that of \textbf{Perfect-CSI \& OAMP-DD}, which relies on the perfect CSI. As expected, \textbf{Perfect-CSI \& SDE-DD} attains the best performance.

\begin{figure}[!h]
\centering
	\captionsetup{font={footnotesize}}
	\hspace{-6mm}
	\subfigure[NMSE]{
		\includegraphics[width=.25\textwidth,keepaspectratio]{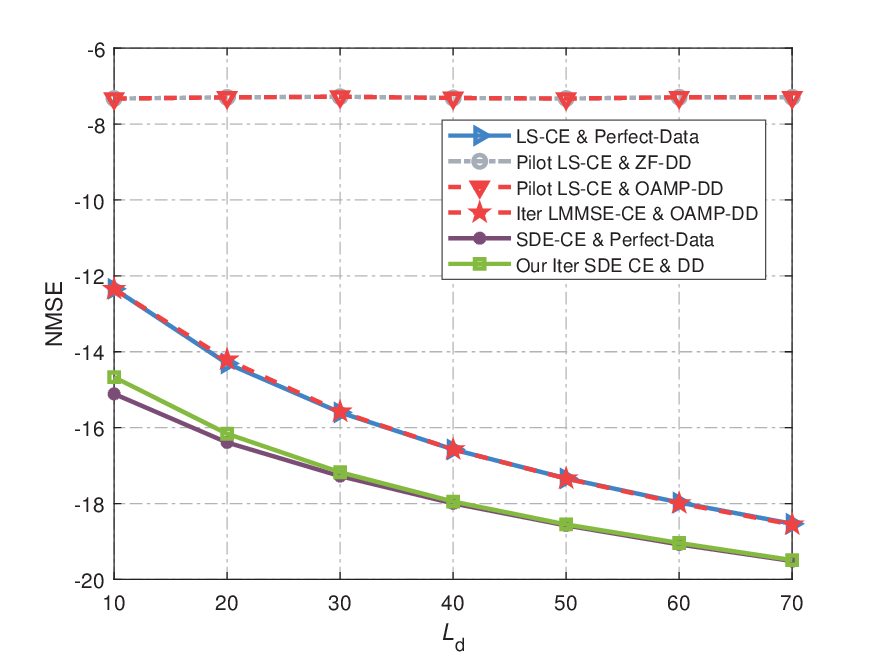}
		\label{fig_jcedd_3a}
	}
	\hspace{-6mm}
	\subfigure[BER]{
		\includegraphics[width=.25\textwidth,keepaspectratio]{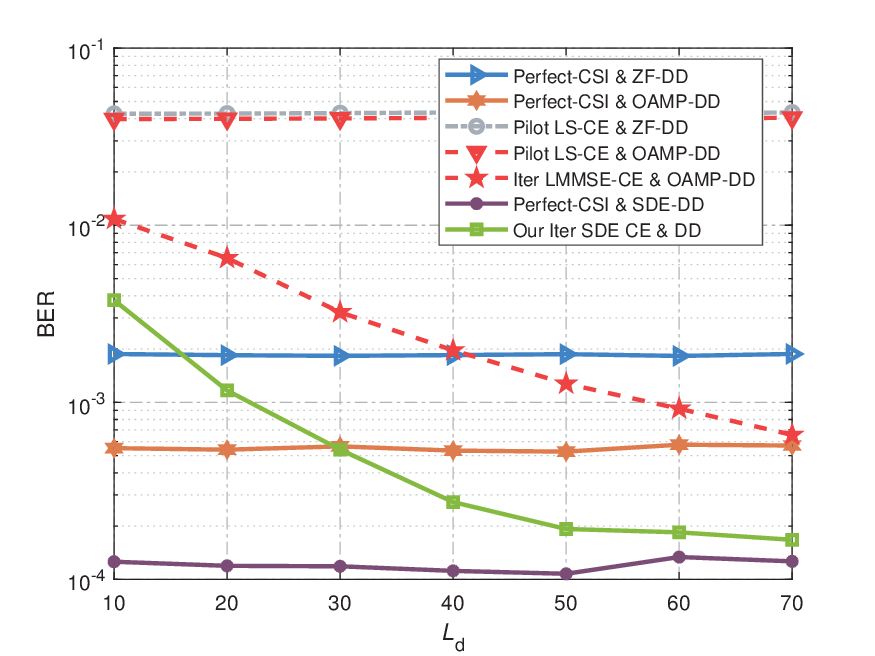}
		\label{fig_jcedd_3b}
	}
	\vspace*{-2mm}
\caption{NMSE and BER performance with respect to the data symbol length $L_d$ for different CE and DD algorithms.}
\label{fig_jcedd_3} 
\vspace*{-1mm}
\end{figure}

Fig.~\ref{fig_jcedd_3} depicts the performance curves of CE and DD across various schemes as the functions of data length $L_d$. In the simulation, the pilot length for each UE is fixed at $L_p\! =\! 15$, with the received $\text{SNR}\! =\! 10$\,dB at the BS. As can be seen from Fig.~\ref{fig_jcedd_3a}, the iterative algorithm improves CE performance as data length increases. For $L_d\! \geq\! 30$, the NMSE of \textbf{Our Iter SDE CE \& DD} matches that of \textbf{SDE-CE \& Perfect-Data}, offering an NMSE gain of 1 to 2 dB over \textbf{Iter LMMSE-CE \& OAMP-DD}.  As depicted in Fig.~\ref{fig_jcedd_3b}, increasing $L_d$ improves the accuracy of CE, thereby enhancing the BER performance of \textbf{Iter LMMSE-CE \& OAMP-DD} and \textbf{Our Iter SDE CE \& DD}. Moreover, when $L_d\! =\! 70$, the BER of \textbf{Iter LMMSE-CE \& OAMP-DD} approximates that of \textbf{Perfect-CSI \& OAMP-DD}, while \textbf{Our Iter SDE CE \& DD} surpasses \textbf{Perfect-CSI \& OAMP-DD}, in terms of BER.

\begin{figure}[!t]
\centering
	\captionsetup{font={footnotesize}}
	\hspace{-6mm}
	\subfigure[NMSE]{
		\includegraphics[width=.25\textwidth,keepaspectratio]{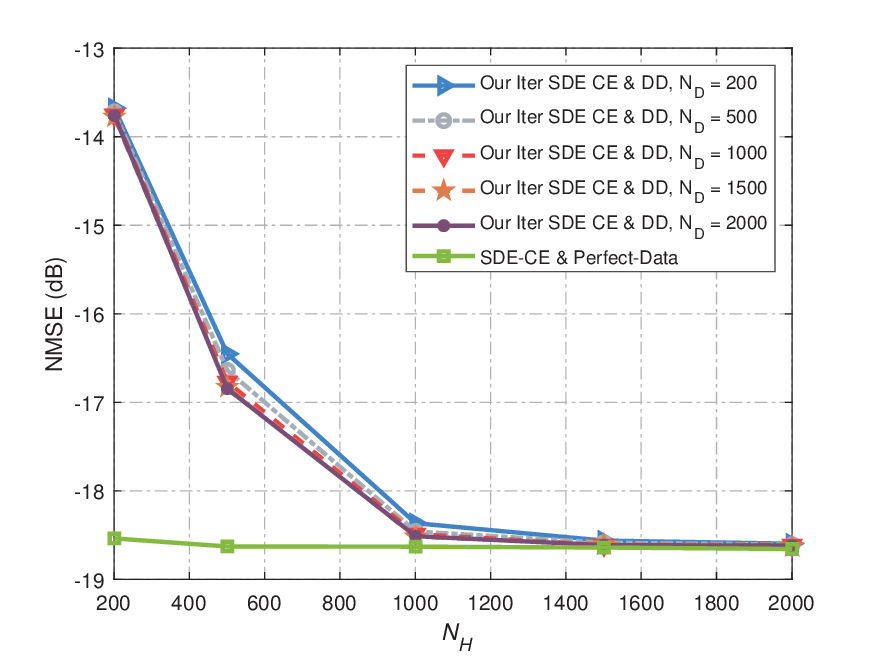}
		\label{fig_jcedd_4a}
	}
	\hspace{-6mm}
	\subfigure[BER]{
		\includegraphics[width=.25\textwidth,keepaspectratio]{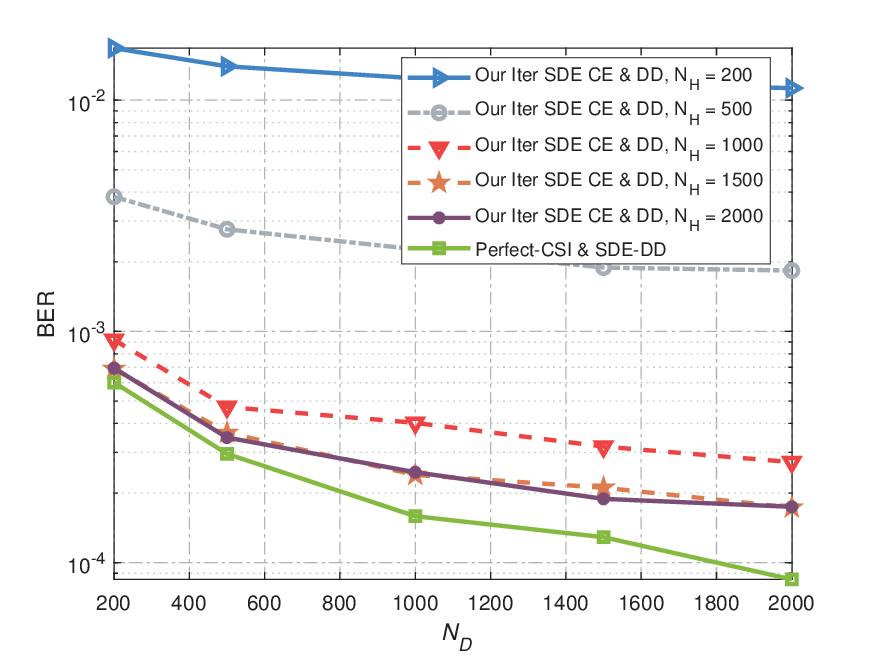}
		\label{fig_jcedd_4b}
	}
	\vspace*{-2mm}
\caption{NMSE and BER performance with respect to the step numbers $N_H$ and $N_X$ of our Iter SDE CE \& DD algorithm.}
\label{fig_jcedd_4} 
\vspace*{-5mm}
\end{figure}

Fig.~\ref{fig_jcedd_4} compares the performance of CE and DD of \textbf{Our Iter SDE CE \& DD} scheme as the function of the maximum algorithm iteration numbers $N_H$ and $N_X$. In the simulation, each UE's pilot length is set to $L_p\! =\! 15$ and the data length is $L_d\! =\! 50$, with the received SNR fixed at 10\,dB. For the predefined noise diffusion functions $\sigma(t)$ and $\tau(t)$, the interval $t\! \in\! [0,\, 1]$ can be divided into segments with different values of $N_X$ and $N_H$ to control the noise variance gap between adjacent steps, thereby balancing generation speed and performance. As shown in Fig.~\ref{fig_jcedd_4a}, as the channel sampling steps $N_H$ increase, the NMSE performance improves gradually due to a more elaborate denoising process, resulting in more accurate SDE solutions. When $N_H\! =\! 1000$, the NMSE of \textbf{Our Iter SDE CE \& DD} approximates that of \textbf{SDE-CE \& Perfect-Data}. Moreover, under the same $N_H$, CE performance remains similar for different $N_D$, possibly due to SDE-based CE not being sensitive to DD errors. Additionally, Fig.~\ref{fig_jcedd_4b} shows that as the data sampling steps $N_D$ increase, the BER performance of DD gradually improves. However, the BER curves vary for different $N_H$, indicating that for DD problems, SDE-based DD methods are more sensitive to CE errors.
\begin{figure}[!t]
	\centering
	\hspace{-6mm}
	\captionsetup{font={footnotesize}}
	\subfigure[]{\label{fig.rl_iter_lp_nmse}%
		\includegraphics[width=.25\textwidth]{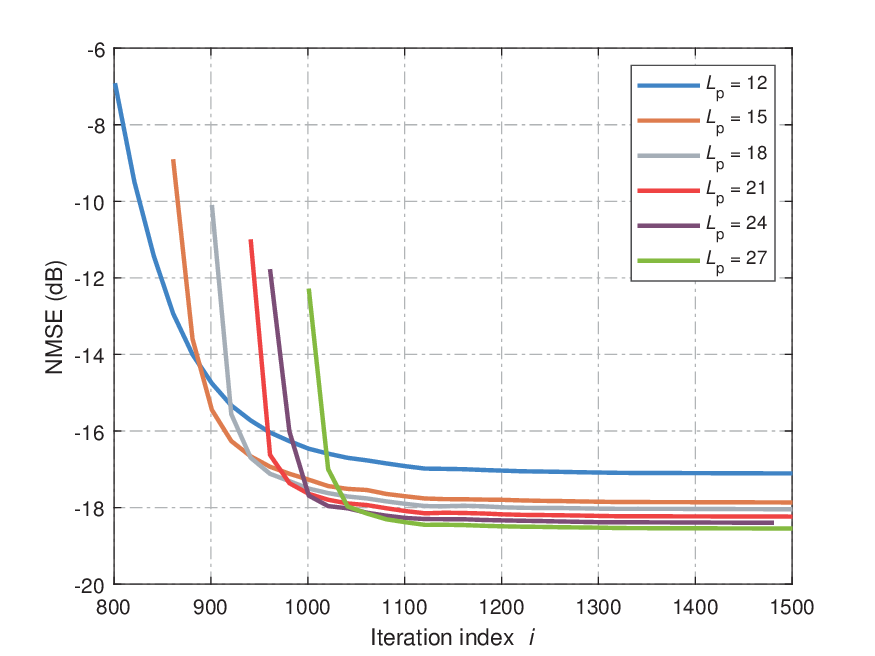}}
	\hspace{-5mm}
	\subfigure[]{\label{fig.rl_iter_snr_ber}%
		\includegraphics[width=.25\textwidth]{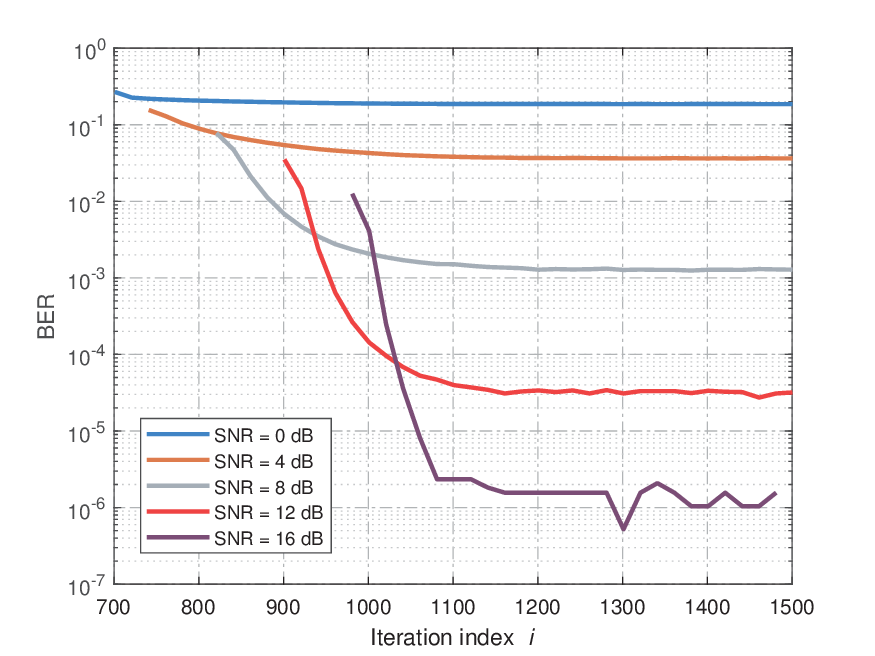}}
	\caption{Convergence performance of (a) NMSE with varying pilot lengths $L_p$ and (b) BER under different SNRs, both as a function of the iteration index $i$.}
	\label{fig.rl_iter_NH}
	\vspace{-2mm}
\end{figure}

Fig. \ref{fig.rl_iter_NH}(a) illustrates the NMSE performance curve  of \textbf{Our Iter SDE CE \& DD} relative to the iteration index $i$ with varying pilot lengths $L_p$. The maximum number of  iterations is set to $N_H = 1500$ ($i \le 1500$). The simulation results indicate that as the pilot length $L_p$ increases, the convergence iteration count for the proposed algorithm decreases from approximately 400 iterations at $L_p=12$ to about 100 iterations at $L_p=27$. Additionally, the final converged NMSE improves from $-17.1$ dB to $-18.5$ dB. Likewise, Fig. \ref{fig.rl_iter_NH}(b) demonstrates that with different SNRs, both the convergence speed and the final DD outcome of our algorithm improve as the SNR increases. Despite employing a maximum of 1500 iterations, the simulation results reveal that under favorable conditions, such as $L_p=27$ and $\text{SNR}=16$ dB, the proposed scheme requires only about 50 to 100 iterations to achieve satisfactory results.

\begin{table}[!h] 
	\centering
	\captionsetup{font=footnotesize}
	\caption{Computation latency of Algorithm 2 at different pilot lengths}  
	\label{rl_tab_latency_vs_lp} 
		\resizebox{1\columnwidth}{!}{
			\renewcommand\arraystretch{1.5}{
				\begin{tabular}{|c|c|c|c|c|c|c|c|}
					\hline
					Metrics    & Methods    & $L_p= 12$ & $L_p = 15$ & $L_p = 18$ & $L_p = 21$ & $L_p = 24$ & $L_p = 27$ \\ \hline
					\multirow{3}{*}{Latency {[}s{]}} & w/ LMMSE init      & 2.32      & 2.14       & 1.99       & 1.87       & 1.76       & 1.70       \\ \cline{2-8} 
					& w/o LMMSE init     & 4.89      & 4.88       & 4.88       & 4.87       & 4.86     & 4.88       \\ \cline{2-8} 
					& relative time cost & 47.41\%   & 43.86\%    & 40.84\%    & 38.42\%    & 36.14\%    & 34.79\%    \\ \hline
	\end{tabular}}}
\end{table}
\vspace{-5mm}
\begin{table}[!h]
	\centering
		\captionsetup{font=footnotesize}
		\caption{Computation latency of Algorithm 2 at different SNRs}  
		\label{rl_tab_latency_vs_snr}
		\resizebox{1\columnwidth}{!}{
			\renewcommand\arraystretch{1.5}{
				\begin{tabular}{|c|c|c|c|c|c|c|}
					\hline
					Metrics                           & Methods            & $\text{SNR}= 0$ dB & $\text{SNR}= 4$ dB & $\text{SNR}= 8$ dB & $\text{SNR}= 12$ dB & $\text{SNR}= 16$ dB \\ \hline
					\multirow{3}{*}{Latency {[}s{]}} & w/ LMMSE init      & 2.62            & 2.39            & 2.16            & 1.84             & 1.44             \\ \cline{2-7} 
					& w/o LMMSE init     & 4.87            & 4.89            & 4.88          & 4.89             & 4.88             \\ \cline{2-7} 
					& relative time cost & 53.76\%         & 48.82\%         & 44.37\%         & 37.61\%          & 29.52\%          \\ \hline
	\end{tabular}}}
\end{table}	

In Tab. \ref{rl_tab_latency_vs_lp}, we present the average computation latency for one single realization of \textbf{Our Iter SDE CE \& DD} algorithm, both with and without LMMSE initialization, across various pilot lengths. Simulations were performed using an NVIDIA GeForce GTX 4090 GPU and an AMD EPYC 7J13 CPU. As the pilot length $L_p$ increases, the computational latency of the method with LMMSE initialization gradually decreases, whereas the latency without LMMSE initialization remains relatively constant. This occurs because a longer pilot results in more accurate initial channel estimation, which corresponds to a smaller diffusion noise-level in the iteration process, resulting in fewer required iteration steps. Similarly, Tab. \ref{rl_tab_latency_vs_snr} illustrates the differences in computational latency across various SNRs when LMMSE initialization is applied or not. Notably, our approach requires only 29\%-53\% of the original scheme's complexity, underscoring the effectiveness of the LMMSE initialization strategy.

\begin{figure}[!h]
	\centering
	\captionsetup{font={footnotesize}}
	\includegraphics[width=0.9\columnwidth]{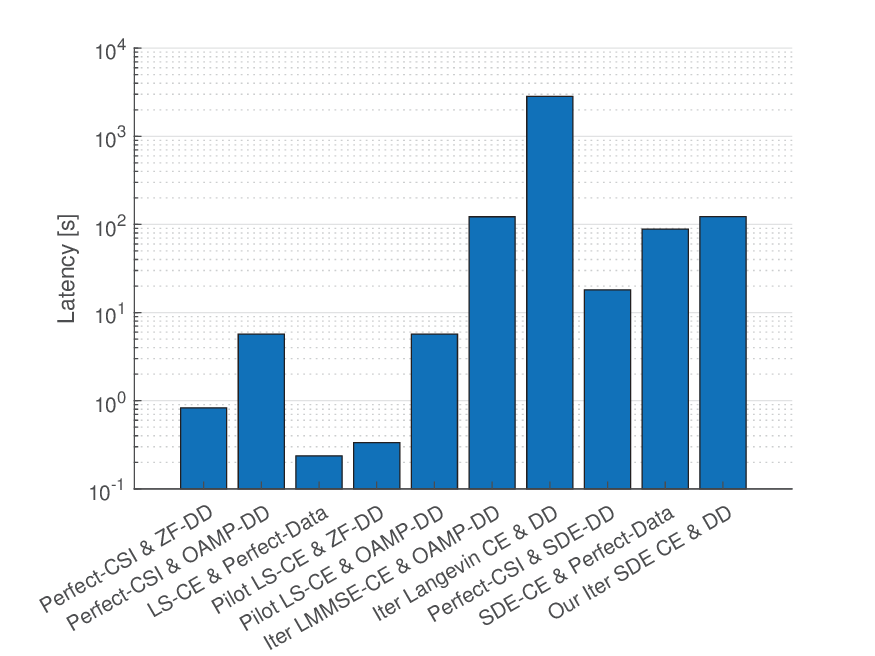}
	\caption{Computation latency of different CE and DD schemes.}
	\label{fig_jcedd_5} 
	\vspace{-1mm}
\end{figure}

Fig.~\ref{fig_jcedd_5} compares the computational latency of various methods for processing a batch size of 64 in CE and DD tasks. Due to its asynchronous update mechanism and LMMSE initialization, \textbf{Our Iter SDE CE \& DD} algorithm demonstrates significantly lower computational latency compared to \textbf{Iter Langevin CE \& DD} algorithm. Furthermore, the latency of \textbf{Our Iter SDE CE \& DD} is comparable to that of \textbf{Iter LMMSE-CE \& OAMP-DD}, while the previous simulations have shown it outperforming \textbf{Iter LMMSE-CE \& OAMP-DD} in both CE and DD. Although the latency of iterative CE and DD algorithms is higher than that of two-stage algorithms, such as \textbf{Pilot LS-CE \& ZF-DD} and \textbf{Pilot LS-CE \& OAMP-DD}, existing researches indicate that diffusion-based algorithms can reduce the number of steps from thousands to tens, or even to as few as 1 to 2 steps, through acceleration methods such as higher-order algorithms \cite{dpmv3_23nips} and consistency models \cite{song-23icml}. Therefore, accelerating diffusion algorithms still holds significant potential for addressing CE and DD problems.

\section{Conclusions}\label{S7}

We have proposed techniques for addressing the challenges of AUD and JCEDD in massive random access, where standard deep learning often struggles to generalize across various problem dimensions using a single model. Our VPL-AUDNet architecture has been designed for variable pilot length transmission, enabling a single model to adapt to different pilot lengths and antennas, thereby achieving superior AUD performance in both 3GPP and Rayleigh channels. Furthermore, we have introduced a generative diffusion models-driven JCEDD scheme using PC samplers, which significantly enhances CE and DD performance through an asynchronous iterative algorithm. These proposed schemes have demonstrated the potential of utilizing deep learning and generative models to address signal detection and estimation challenges in massive random access. A future research direction of interest is to consider accelerating generative model methods to further reduce inference latency.

\begin{appendices}
	\section{Structure of the Heterogeneous Transformer Network}\label{apd1}
	\begin{figure*}[t!]
		\centering
		\subfigure[]{\label{fig.aud_ht_a}%
			\includegraphics[width=.52\textwidth]{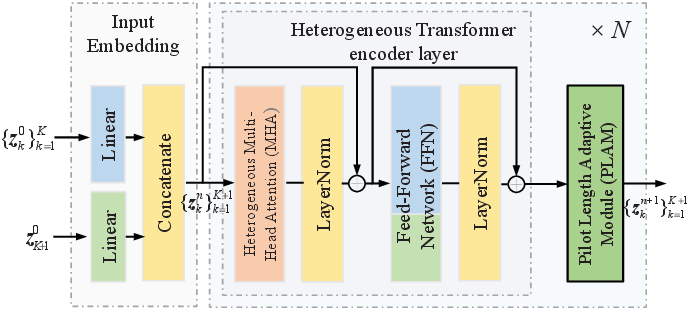}}
		\subfigure[]{\label{fig.aud_ht_b}%
			\includegraphics[width=.29\textwidth]{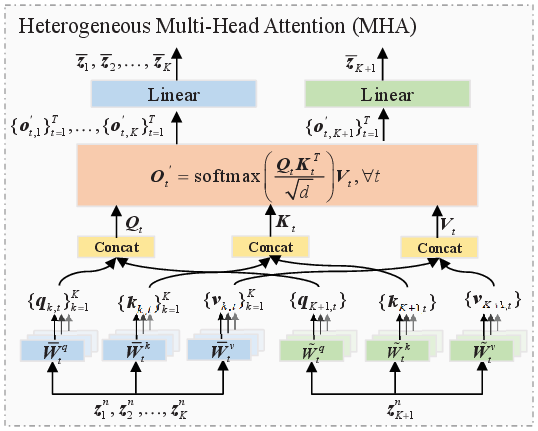}}
		\captionsetup{font={footnotesize}}
		\vspace{-3mm}
		\caption{Block diagrams of (a) the heterogeneous Transformer network and the PLAM in the intermediate layer of the proposed VPL-AUDNet, and (b) the heterogeneous MHA within the heterogeneous Transformer encoder layer.}
		\label{fig.aud}
	\end{figure*}
	Fig. \ref{fig.aud_ht_a} illustrates the structure of the intermediate layer in the proposed VPL-AUDNet network. First, for the input pilot feature sequences $\{\bm{z}_k^0\}_{k=0}^{K}$, and the received signal feature sequence $\bm{z}_{K+1}^0$, two distinct linear layers are employed for linear embedding. This process results in the sequences, $\{\bm{z}_{k}^{1}\}_{k=1}^{K}$, with dimensions $(B, K+1, d)$, where $B$ is the batch size, $K+1$ is the sequence length, and $d$ is the feature dimension.  Subsequently, the sequence undergoes processing through $N$ alternating heterogeneous Transformer encoder layers and PLAM networks, and finally produce $\{\bm{z}_{k}^{N+1}\}_{k=1}^{K}$ for the output layer of the VPL-AUDNet. Fig.  \ref{fig.aud_ht_a} also demonstrates that the unique aspect of the heterogeneous Transformer network,  in contrast to the traditional Transformer, is its use of distinct linear matrix weights for input embedding, MHA, and FFN components.
	
	Fig. \ref{fig.aud_ht_b} details the computation process of heterogeneous MHA.  In the traditional Transformer, the query matrices $\bm{Q}_t$ for the $t$-th head are calculated using the same weight matrices ${\bm{W}}_{t}^{q}$ across all $K+1$ input sequences, that is, $\bm{Q}_t = {\bm{W}}_{t}^{q}\bm{z}_{k}$, where $k \in \{1, \ldots, K+1\}$. 
	In contrast, we employ two distinct sets of matrices, ${\bar{\bm{W}}_{t}}^{q}$ and ${\tilde{\bm{W}}_{t}}^{q}$, to separately derive the query results for the pilot sequence and the received signal sequence, as illustrated at the bottom of Fig. \ref{fig.aud_ht_b}: $\bm{q}_{k,t} = {\bar{\bm{W}}}_{t}^{q}\bm{z}_{k}$, for $k \in \{1, \ldots, K\}$, and $\bm{q}_{K+1,t} = {\tilde{\bm{W}}}_{t}^{q}\bm{z}_{K+1}$.  
	Subsequently, we concatenate the sequences $\{\bm{q}_{k,t}\}_{k=1}^{K}$ and $\bm{q}_{K+1,t}$ into $\bm{Q}_{t} = \left[\bm{q}_{1,t}, \ldots, \bm{q}_{K,t}, \bm{q}_{K+1,t}\right]$ for further calculations. A similar procedure is applied to compute the key matrices $\bm{K}_t$ and value matrices $\bm{V}_t$. Finally, after matrix multiplication and the softmax operation, the output across different heads is divided into the first $K$ sequences and the $(K+1)$-th sequence. Different linear layers are utilized for multi-head fusion, resulting in $\{\bar{\bm{z}}_k\}_{k=1}^{K+1}$, which are then forwarded to the subsequent module.
\end{appendices}

\end{document}